\documentclass{article}

\usepackage{PRIMEarxiv}

\usepackage[utf8]{inputenc} % allow utf-8 input
\usepackage[T1]{fontenc}    % use 8-bit T1 fonts
\usepackage{hyperref}       % hyperlinks
\usepackage{url}            % simple URL typesetting
\usepackage{booktabs}       % professional-quality tables
\usepackage{amsfonts}       % blackboard math symbols
\usepackage{nicefrac}       % compact symbols for 1/2, etc.
\usepackage{microtype}      % microtypography
\usepackage{lipsum}
\usepackage{fancyhdr}       % header
\usepackage{graphicx}       % graphics
\graphicspath{{media/}}     % organize your images and other figures under media/ folder
\usepackage{stfloats}
\usepackage{subcaption}
\usepackage{xspace}
\usepackage{amssymb}

\usepackage{tikz}

\makeatletter
\newcommand*{\radiobutton}{%
  \@ifstar{\@radiobutton0}{\@radiobutton1}\xspace%
}
\newcommand*{\@radiobutton}[1]{%
  \begin{tikzpicture}
    \pgfmathsetlengthmacro\radius{height("X")/2}
    \draw[radius=\radius] circle;
    \ifcase#1 \fill[radius=.6*\radius] circle;\fi
  \end{tikzpicture}%
}

\makeatother

\newcommand{\eg}{\textit{e.g.,}\xspace}
\newcommand{\etc}{\textit{etc.}\xspace}

\newcommand{\quotes}[1]{``#1''}

\title{Evaluation of Virtual Reality Interaction Techniques: the case of 3D Graph}

\author{
  Nicola Capece, Ugo Erra, Monica Gruosso \\
  University of Basilicata \\
  Potenza, Italy\\
  \texttt{\{nicola.capece, ugo.erra, monica.gruosso\}@unibas.it} \\
   \And
  Delfina Malandrino \\
  University of Salerno \\
  Salerno, Italy\\
  \texttt{dmalandrino@unisa.it} \\
 \and
  Max M. North \\
  Kennesaw State University \\
  Marietta, USA\\
  \texttt{dr.max.north@gmail.com} \\
}

\begin{document}
\maketitle

\begin{abstract}
The virtual reality (VR) and human-computer interaction (HCI) combination has radically changed the way users approach a virtual environment, increasing the feeling of VR immersion, and improving the user experience and usability. The evolution of these two technologies led to the focus on VR locomotion and interaction. Locomotion is generally controller-based, but today hand gesture recognition methods were also used for this purpose. However, hand gestures can be stressful for the user who has to keep the gesture activation for a long time to ensure locomotion, especially continuously. Likewise, in Head Mounted Display (HMD)-based virtual environment or Spherical-based system, the use of classic controllers for the 3D scene interaction could be unnatural for the user compared to using hand gestures such \eg pinching to grab 3D objects. To address these issues, we propose a user study comparing the use of the classic controllers (six-degree-of-freedom (6-DOF) or trackballs) in HMD and spherical-based systems, and the hand tracking and gestures in both VR immersive modes. In particular, we focused on the possible differences between spherical-based systems and HMD in terms of the level of immersion perceived by the user, the mode of user interaction (controller and hands), on the reaction of users concerning usefulness, easiness, and behavioral intention to use.  
\end{abstract}

% keywords can be removed

\section{Introduction}\label{sec:intro}
The development of interaction and locomotion approaches represents the new challenge of VR technologies. Indeed, traditional approaches such as controllers~\cite{LI2019, McMahan} have been overcome through the combination of VR and HCI methods. In this direction, more and more approaches were proposed~\cite{pfeuffer2017, Piumsomboon2017} considering also locomotion~\cite{Metsis2017, caggianese2020, Pyo2018} studies useful to increase the level of user immersion in the virtual environment. 
Hand tracking and gesture recognition represent one of the most popular novelties of VR with HCI applications~\cite{Capece2020, Szabo2019, Lee2015, ragazzoni2018}. Specific devices such as omnidirectional treadmill (ODT)~\cite{HOOKS2020100352}, hand tracking methods onboard of HMDs~\cite{Voigt-Antons2020, Hillmann2019} were affirmed and consolidated, allowing a significant increase in the user's sense of immersion in VR. However, ODT, for example, requires ad-hoc hardware which needs effective technology as support, such as haptic and harness tools, to ensure a low-stress and comfortable low-stress user experience.   
Hand tracking and gesture recognition-based approaches are affected by the tired arm and muscle fatigue which is called also \quotes{gorilla arm effect}. This is due to the necessity to keep for long time hands visible to the tracking devices, such as Leap Motion, Dept Cameras, simple Webcam, \etc In particular, this effect is more evident when the hand gestures are used for the user's locomotion continuous because he has to keep the arm still for a long time, consequently increasing stress and fatigue~\cite{caggianese2020, wiedemann2017experiment}. 
These problems are further accentuated in spatial data exploration contexts~\cite{Hayatpur2020} where there is a need to understand how data are structured and related. Indeed, a new research field called Immersive Analytics~\cite{Bach2016} has recently emerged with the aim of understanding how visualization and interaction technologies can assist data scientists in the data analysis through the user immersion in the data itself~\cite{drogemuller2017vrige}. The relationships between big data are often visualized through graph structures to understand their intrinsic features. Indeed, the visualization of large graphs has been empirically proven to be a tool that greatly improves the ability to understand the relationships between objects~\cite{Muelder2013, ERRA201813}. In this context, visualizing data on traditional displays and interacting with them through standard devices such as keyboard and mouse is becoming a trend cumbersome~\cite{Drogemuller2018}. To address these issues, we propose a comparative study among different approaches for user interaction, locomotion, and visualization in VR. On the other hand, in this paper, we propose an evaluation study involving $60$ participants divided into $4$ groups of $15$ users. We asked each group to interact and locomote in VR scene using different devices and methodologies: the first group used the HMD HTC Vive for VR immersion and its 6-DOF controllers to interact with virtual objects and locomote within the VR scene ($HTCVive\_2Controllers$); we asked the second group to visualize the VR scene using the same HMD and a multimodal approach~\cite{capece2021} using the 6-DOF controller to locomote around the VR scene and hand tracking based on a Leap Motion device placed on top of HMD to interact with the virtual objects ($HTCVive\_Controller\_LeapMotion$); third group instead used an Immersive Visualization Environment (IVE) (spherical-based virtual environment), discussed in Section~\ref{sec:IVE} and two trackballs to interact with 3D objects and locomote in the VR scene ($IVE\_2Trackballs$); in the last group we asked to use IVE for VR visualization, a trackball for locomotion and hand tracking through Leap Motion device placed on the user's forehead ($IVE\_Trackball\_LeapMotion$).
The proposed use case is represented by a virtual 3D scene in which users have to explore a 3D graph and interact with the nodes or with the entire graph itself. 
The results of our comparative study show that the spherical systems for VR purposes are promising and competitive with respect to the HMD systems, and the interaction and locomotion with two controllers still represent the optimal choice for HMD-based environments. However, the multimodal approach based on a single trackball and Leap Motion device provides interesting results in terms of usability and accuracy within the IVE-based environment.

The remainder of this paper is structured as follows: an overview of the state-of-the-art is reported in Section~\ref{sec:related}; some details on the background and methods used in our approach can be found in Section~\ref{sec:back}; a description of the 3D graph and details on VR scene is reported in Section~\ref{sec:case}; the evaluation study and details on the experiment can be found in Section~\ref{sec:evaluationstudy}; results and discussion are reported in Section~\ref{sec:results}; finally, conclusions and future directions of our work are reported in Section~\ref{sec:conc}.

\section{Related Work}\label{sec:related}
Interaction with virtual objects and Locomotion in a virtual scene requires a high user effort~\cite{Iqbal2021, Hincapi2014, Jang2017, Sarupuri2017, DeChiara2007203}, especially through freehand locomotion techniques with continuous controlling, where the user has to keep the hands in the same position for all movement time~\cite{Zhang2017, caggianese2020}. An empirical evaluation study was proposed from~\cite{caggianese2020} about a comparison among freehand-steering locomotion techniques and a controller-based approach. The focus was to evaluate the efficiency, effectiveness, and user preference in continuously controlling the locomotion direction using controller-steering, palm-steering, index-steering, and gaze-steering techniques. Eye gaze is also used in a VR interaction method proposed in \cite{pfeuffer2017} to select virtual objects and the manipulation was ensured through freehand gestures. In particular, the authors proposed gestures as two-handled scaling or pinch-to-select to the virtual objects observed by the user. \cite{Zhang2017} proposed a double-hand gestures locomotion technique, where the right thumb was used to turn left or right and the left palm was used to control the backward and forward movements. Another interesting approach based on hand gestures was proposed from ~\cite{Satriadi2019}, which is focused on AR intangible digital map navigation. The authors conducted a study exploring the handedness effect and input mapping defining two techniques to transit between positions smoothly. They reported that input-mapping transitions could increase performance and reduce arms fatigue. One of the freehand input limitations is the different operations that can be invoked concerning the user's intention. This effect was investigated from~\cite{Chen2020}, who proposed an experimental analysis to evaluate some techniques to disambiguate the effect of the freehand manipulations in VR. The study was conducted by comparing the hand gaze, speech, and foot tap input methods, putting together three-timing (before, during, and after an interaction) in which they provide settings to resolve the ambiguity.  
To allows the arm fatigue, ~\cite{Iqbal2021} proposed a combination of feasible techniques called \quotes{ProxyHand} and \quotes{StickHand}. The first enabled the user to interact with the VR scene through his arm in a comfortable position through a 3D-spatial offset between his real hands and the virtual representation. The second is used where the \quotes{ProxyHand} was not suitable. \cite{Sarupuri2017} proposed a VR low-fatigue hand-controller-based travel technique for a limited physical space. Through a single controller, the average of hand and controllers, or the head, the user can choose the travel orientation.
A comparison between VR systems and traditional mouse-keyboard and joypad configurations was proposed from~\cite{ErraMP19}. The authors proposed an empirical evaluation study about the interaction and locomotion with 3D graphs using the Oculus Rift and Leap Motion, designing a specific natural user interface. They developed a plug-in module for the Gephi software to evaluate the 3D graph interaction through VR and standard monitor combined with keyboard-mouse, joypad, and Leap Motion input devices. In such a study, the authors concluded that the VR based on HMD was more challenging compared with the traditional devices. 
Exploring and interacting with 3D graphs in VR was also investigated in~\cite{Capece2018}, in which we studied the better graph layout which allows the nodes to be positioned to reduce as more as possible entropy levels. 
To the best of our knowledge, there are no specific studies that compare VR visualization systems based on IVE and HMD and at the same time compare different modalities of interactions, hand tracking, gesture recognition based, controller based, and trackball based. 

\section{Background and Methods}\label{sec:back}

\begin{figure*}[ht!]
    \centering
    \includegraphics[width=1.0\textwidth]{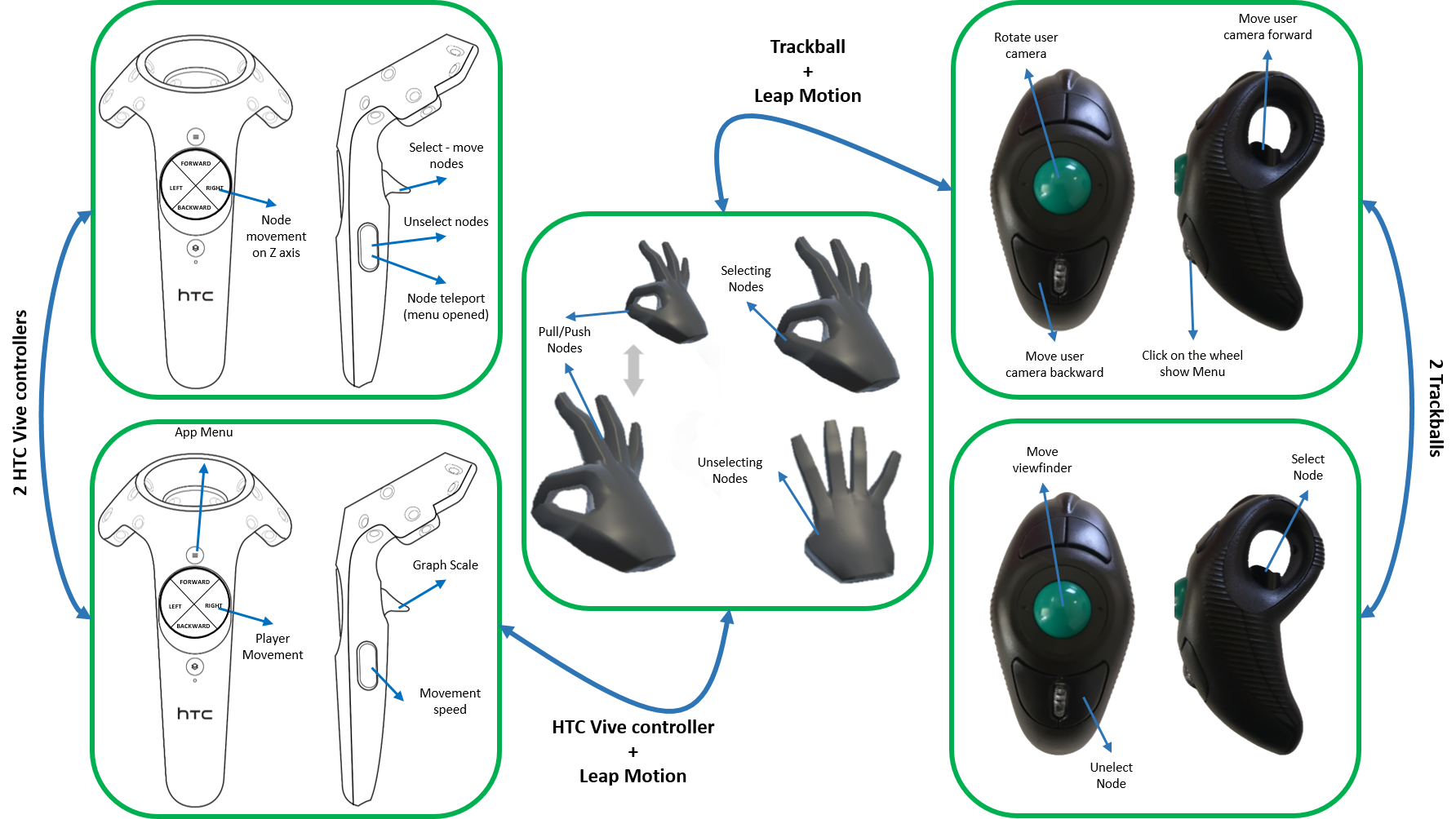}
    \caption{%The scheme of the multimodal approach. On the left, the features associated with the HTC Vive controller helpful to interact with the menu and for the locomotion in the scene. On the right, the gestures associated with hand tracking using the leap motion controller to interact with the scene components (the nodes).
    On the left, are the features associated with the HTC Vive controllers to interact and locomote in the 3D Graph. On the center, the gestures associated with the hand tracking using the Leap Motion controller to interact with the scene components. On the right, the features associated with the trackballs to locomote and interact with the 3D graph. The arrows indicate the $4$ configurations of the input devices based on the experiment they were used.} 
    \label{fig:controller-hand}
\end{figure*}

On the left, are the features associated with the HTC Vive controllers to interact and locomote in the 3D Graph. On the center, the gestures associated with the hand tracking using the Leap Motion controller to interact with the scene components. On the right, the features associated with the trackballs to locomote and interact with the 3D graph. The arrows indicate the $4$ configurations of the input devices based on the experiment they were used

The proposed evaluation study was conducted on a well-designed scene based on the 3D Graph visualization (see Section~\ref{sec:case}). The graph structure used for our purposes was generated and exported using the Gephi~\cite{gephi2009} software, which allows us to define colors, size, node positions, and edges defined through adjacent nodes. The Force Atlas 2~\cite{Jacomy2014} 3D layout was used as graph layout and the overall graph description was exported through a \textit{.gexf} output file. The latter is the standard XML-based Gephi format which can be imported into our 3D scene using a parsing function properly developed. Furthermore, we tested the robustness of our 3D application, generating a random large graph with a growing variable number of edges and nodes. 
The imported \textit{.gexf} file was scanned from a Unity 3D prefab we developed, called \textit{PlacePrefabPoint}. Prefab is a serialized off-the-shelf Unity 3D game object which was previously created and configured. It creates another specific prefab for each node found in the \textit{.gexf} file, tying it with its graphic features such as sphere size, color, position \etc. \textit{PlacePrefabPoint} creates also for each node, an empty adjacency list, which is filled with its connected nodes defined by the edges also reported in the \textit{.gexf} file. The filling operation was performed with another prefab that we called \textit{PlacePrefabEdge}.
Our study case was developed using the unity 3D game engine which allowed us to quickly integrate the Leap Motion SDK for hand gesture recognition, the Steam VR framework to support the HTC Vive HMD and its controllers, and the screen-keyboard-mouse paradigm. In the next subsections we discussed the two VR visualization approaches and for each of them the two interaction methods as reported in Figure~\ref{fig:configurations}.

\subsection{HMD}\label{sub:hdm}
\begin{figure*}[ht!]
    \centering
    \begin{subfigure}{0.475\textwidth}
    \includegraphics[width=1\textwidth]{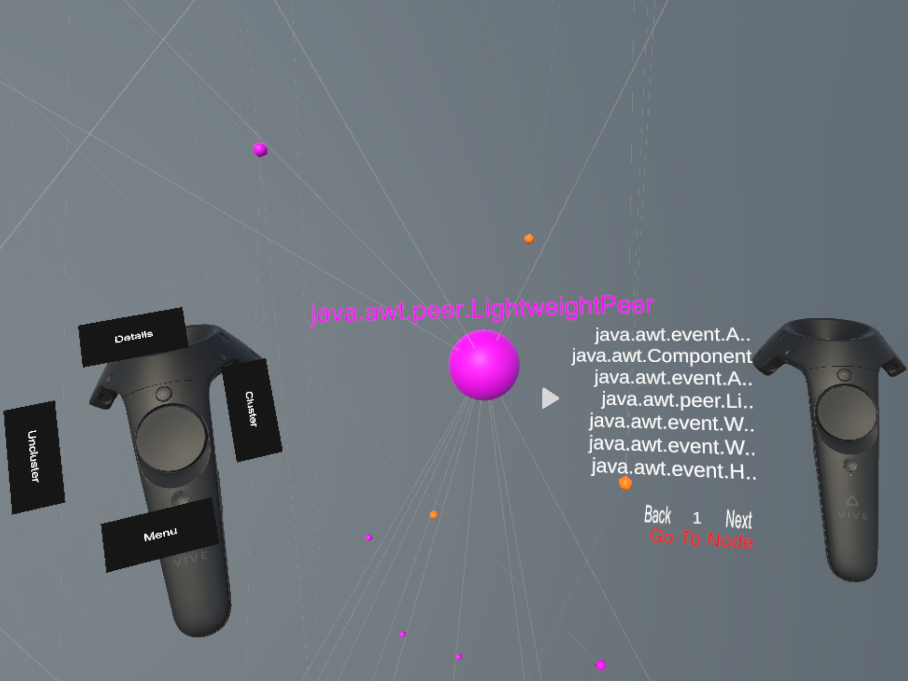}
    \caption{}
    \label{fig:controllers}
    \end{subfigure}
    \hfill
    \begin{subfigure}{0.475\textwidth}
    \includegraphics[width=1\textwidth]{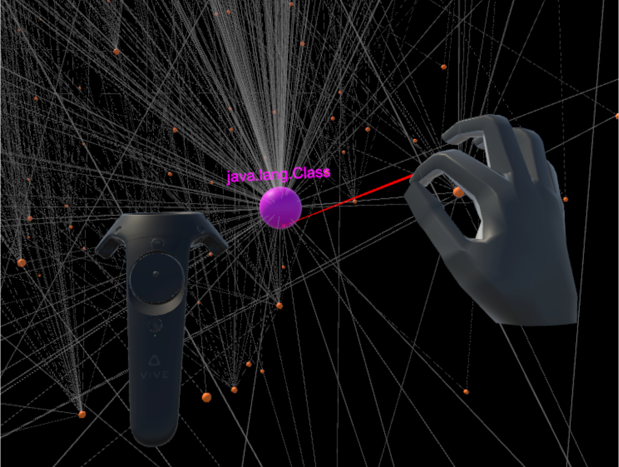}
    \caption{}
    \label{fig:contr_leap}
    \end{subfigure}
    \caption{Image~\ref{fig:controllers} shows the rendering of a 3D graph and the application menu interaction in the $HTCVive\_2Controllers$ experiment. Image~\ref{fig:contr_leap} shows the multi-modal approach based on the controller for user locomotion and the hand tracking and gesture recognition for 3D scene interaction used in the $HTCVive\_Controller\_LeapMotion$ experiment.}
\end{figure*}

To allow the user the VR experience using the HMD, we used HTC Vive (first version) with a $3.6$ inch AMOLED display with $1080 \times 1200$ per eye, $110$ as a field of view degree, and 90Hz as refresh rate. Such HMD allows the user to adapt the interpupillary and lens distances according to his needs, and the allowed range for the user movement is bounded with a defined room scale~\cite{Peer2017} tracked from the base stations. For the $HTCVive\_2Controllers$ experiment, we used an HTC Vive controller (also tracked with base stations) for the locomotion, which is based on a 6-DOF and can be associated with a left or right user hand (see Figure~\ref{fig:controllers}). In particular, the user can move using the controller's trackpad along the horizontal axis: forward, backward, left, and right in flight mode. The vertical movement can be performed through the HMD forward direction by pointing and pressing the forward trackpad button. Furthermore, the locomotion speed can be increased through the Grip Button of the controller (see left part of Figure~\ref{fig:controller-hand}). The same controller allows the user to open and close the application menu. In the same experiment, the second 6-DOF controller is used to interact with the 3D scene graph and the nodes. In $HTCVive\_Controller\_LeapMotion$ experiment, we used the same approach used for $HTCVive\_2Controllers$ for the locomotion and the Leap Motion controller to allows the 3D scene interaction (see Figure~\ref{fig:contr_leap}). This controller is a compact infrared-camera device useful for hand tracking and gesture recognition which can be mounted on an HMD device, allowing the group of this experiment participants to visualize the 3D simulated hands in the virtual scene and faithfully follow the movements of their real hands, increasing the VR immersion feeling.

In both HMD experiments, the buttons of the locomotion controller are used to activate and interact with the application menu. This menu allows the user to change the option setting such as \eg max graph plot scale and max user speed movement of the application. The menu allows also the user to navigate and visualize the node list, select one of them and teleport in front of it. Finally, the menu allows viewing the labels of the selected nodes.

\subsection{Immersive Visualization Environment Experiment}\label{sec:IVE}
IVE,  also known as a  dome-shaped system  (see  Figures  1 and  2) located at Visualization \& Simulation Research Cluster (\url{coles.kennesaw.edu/vsr}).  This state-of-the-art equipment enhances the immersive virtual environment imaging with multiple digital projectors and a large cylindrical screen. The particular immersive system used for the experiment is a spatially immersive visualization and features four digital projectors,  OmniFocus\texttrademark~ Series  500 wide-angle single-lens projection system, and an 8ft by 10ft cylindrical screen with multi-channel visual display, coupled with integrated hardware and software warp and blend technology (see  Figure  2,  a schematic configuration of the IVE system). Numerous variations of IVE are designed and implemented at research and training laboratories in prime industries, such as Boeing, BMW, and Lockheed  Martin, and in branches of the  Department of  Defense,  Army,  Air  Force,  Navy, and National  Security Agency, just to list a few. Such installations have a wide range of applications, such as architectural design, flight simulation, military simulation, scientific visualization, experiential marketing, and industrial simulation~\cite{maxwell2014use}.

\begin{figure}
    \centering
    \includegraphics{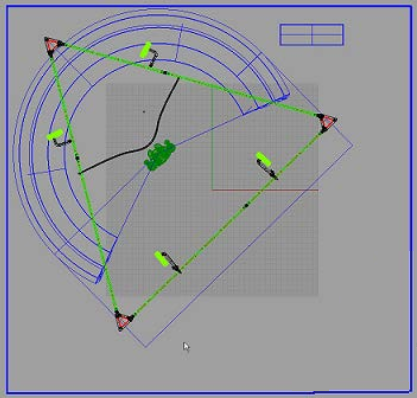}
    \caption{A schematic configuration of the Immersive Visualization Environment system.}
    \label{fig:ive_scheme}
\end{figure}

\begin{figure*}[ht!]
    \centering
    \begin{subfigure}{0.475\textwidth}
    \includegraphics[width=1.0\textwidth]{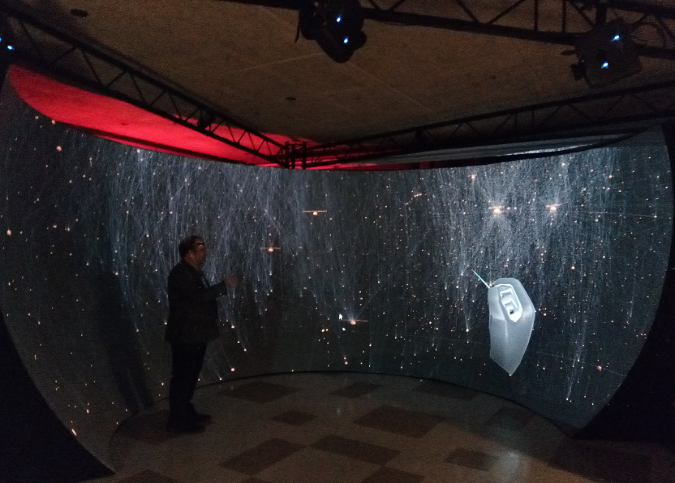}
    \caption{}
    \label{fig:ive_left}
    \end{subfigure}
    \hfill
    \begin{subfigure}{0.475\textwidth}
    \includegraphics[width=1.0\textwidth]{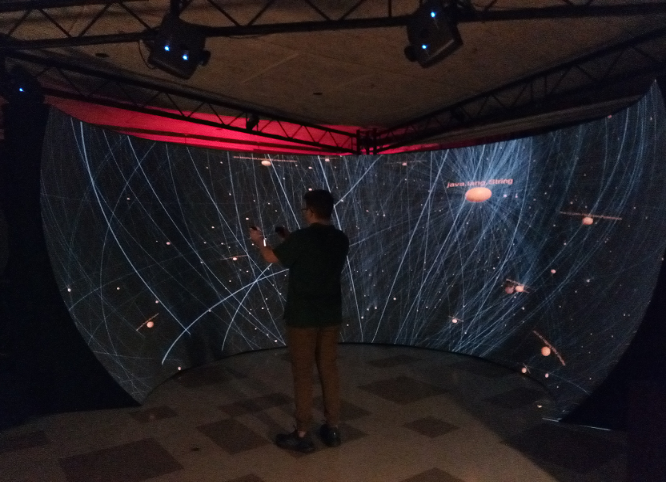}
    \caption{}
    \label{fig:ive_right}
    \end{subfigure}
    \caption{Illustrates (left) a participant controlling leap motion hand and trackball to manipulate 3D graph objects in the Immersive Visualization Environment system; and (right) A subject engaged in the Immersive Visualization Environment depicting a 3D graph using two trackballs.}
\end{figure*}
As IVE can be used with screen-keyboard-mouse paradigm, to allows high user immersion feeling, we used an ergonomic like-mouse trackball. In $IVE\_2Trackballs$ experiment, we configured two trackballs, one for locomotion and the other for interaction. To allow the user orientation during the visualization and locomotion we replaced the HMD head rotational movement with the trackball sphere and buttons for movements. The other trackball is used to allows the user the interaction with the 3D graph and its nodes like $HTCVive\_2Controllers$ (see Figure~\ref{fig:ive_left}). In the last experiment configuration $IVE\_Trackball\_LeapMotion$, we used a multimodal approach with a trackball for the locomotion and the Leap Motion placed with a headband, for the user interaction like $HTCVive\_Controller\_LeapMotion$ (see Figure~\ref{fig:ive_right}). Locomotion trackball allows also the user to interact with the application menu similarly reported in the Sections~\ref{sub:hdm} and~\ref{sec:interaction} by using the wheel button.

\subsection{Interaction}\label{sec:interaction}
User interaction was allowed using the ray casting technique~\cite{ROTH1982109} using $inf$ as collision distance. As shown in Figure~\ref{fig:controller-hand}, in the Leap Motion interaction method, the user has to perform a pinch gesture to select one or more 3D Graph nodes by considering the index and thumb fingers (see Figure~\ref{fig:contr_leap}). A distance range between $2cm$ and $2.5cm$ on index and thumb fingers was considered to activate this gesture. In the $HTCVive\_Controller\_LeapMotion$ experiment, the multiple node selection was allowed by combining the activated pinch gesture with the user gaze direction displayed with a viewfinder~\cite{choe2019} towards the position of the nodes. In this modality, the ray was cast from the position of HMD by following the VR camera direction, and the user has to keep his hand in the HMD field of view because losing the hand tracking deactivates the gesture recognition not allowing him to grab additional nodes. To deselect the nodes, the open hand gesture has to be performed by the user by keeping the fingers except the thumb in the up direction. Indeed, this latter has to e closed toward the palm.
As shown in Figure~\ref{fig:controller-hand}, in the $HTCVive\_2Controllers$, the trigger button of the HMD controller was used to select a target node. If the user keeps the trigger button pressed the ray was cast from the controller head like a lightsaber, by continuously selecting all the nodes it encounters forward of it. Similarly, the user can deselect the already selected nodes using one or both of the controller's grip buttons. In both the HMD interaction methods, the grip button of the locomotion controller can be used to increase the movement speed.
In the $IVE\_Trackball\_LeapMotion$ the viewfinder can be moved using the ball of the locomotion trackball which move also the user camera, and the ray was cast from its position to the forwarding direction (see Figure~\ref{fig:ive_right}). Also, in this case, the hand tracking can be loose if the hand is out of Leap Motion's field of view, and consequently not allowing further node selection. 
In the last $IVE\_2Trackballs$ experiment, the trigger button of the interaction trackball can be used to select the target node by casting the ray from the mouse cursor (represented as a viewfinder) which can be moved through the interaction trackball ball. Also in this case, by keeping the trigger button pressed, the user can select encounters forward nodes and the down buttons to deselect the same nodes. In both the trackball interaction methods, the movement speed can be changed through the application settings.
To provide user feedback, we change their colors to purple when they are selected. Furthermore, we change from green to yellow the color of the ray from when it hit an unselected node and to the red color when it hit an already selected node. When the selected nodes were further selected (see Figure~\ref{fig:controller-hand}, the user can move these nodes and place them in other scene positions by following the viewfinder until they are released. The graph can be also scaled using the trigger button of the HMD controller and moving the tracked hand or the other controller away from each other, or an application setting in the trackball experiments.

\section{3D Graph Scene}\label{sec:case}
\begin{figure}[ht!]
    \centering
    \includegraphics[width=1\textwidth]{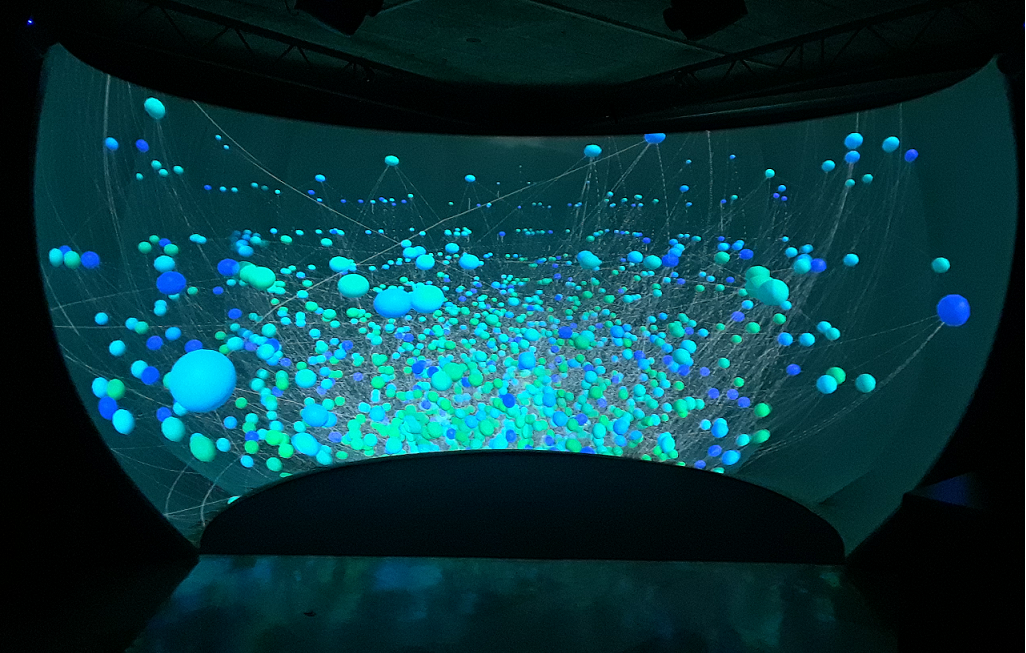}
    \caption{An example of a 3D graph in IVE. As can be seen, the nodes are represented as spheres of which colors, sizes, positions, and edges are retrieved from the \textit{.gexf} file; rendered lines represent edges.
    }
    \label{fig:3Dgraph}
\end{figure}

The reasons behind the proposed virtual scene are motivated by the high user effort required to navigate and interact with the 3D graph. Indeed, the user's ability to move and interact with the scene is severely tested by the representation of complex data structures with a high level of interconnection. The 3D graph is a floating virtual scene and the user has to move in flying mode following all possible directions. As shown in Figure~\ref{fig:3Dgraph}, the graph nodes are represented through basic polygonal spheres~\cite{Capece2018} which allows the user interaction and edges through rendered lines. In this way the user interaction with the edges cannot be carried out since it is a useless feature in our application context and since the use of a non-polygonal representation reduces the complexity of the scene in terms of computing resources and allows to visualize more nodes.

\begin{figure*}[ht!]
    \centering
    \includegraphics[width=1\textwidth]{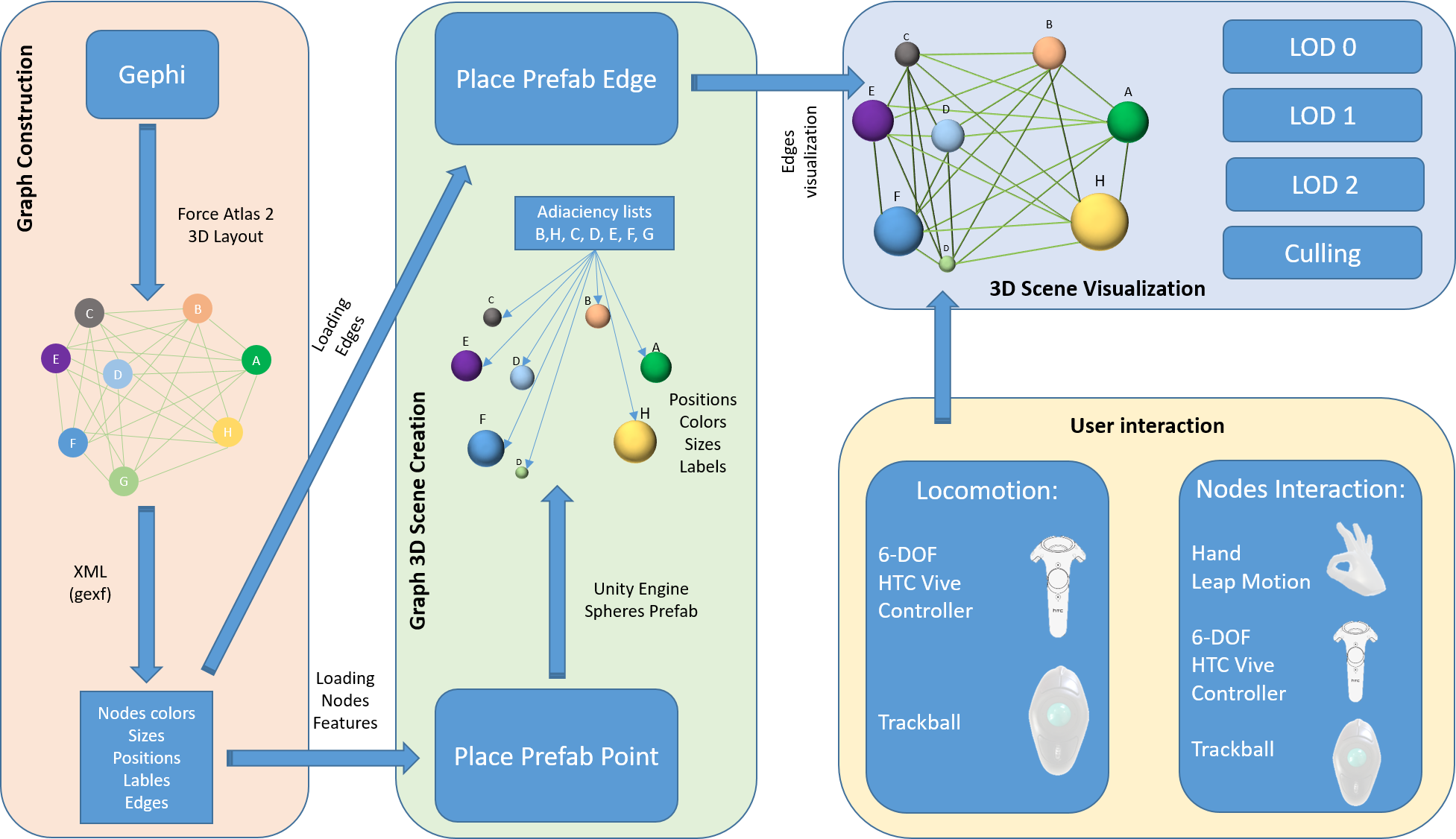}
    \caption{The application architecture design. The pink block shows the features of the graph defined in the \textit{.gexf} XML-based document. The green block shows the 3D graph building using Unity 3D and the two developed prefabs. The blue block shows the 3D scene optimizations, and finally, the yellow block shows the techniques for user locomotion and interaction.}
    \label{fig:workflow}
\end{figure*}

To allow the user the highest level of performance we have developed the scene with a set of tricks. In particular, we noticed the worst rendering time when the user is placed in the 3D graph center or all nodes are visible in the VR camera field of view. As too low render time can lead to motion sickness and general user discomfort~\cite{zhang2020}, we address this problem by introducing three Levels Of Detail (LOD) and an occlusion culling level~\cite{hey2001occlusion} (see Figure~\ref{fig:workflow}). Thus, the 3D nodes' geometric complexity was automatically adapted based on their distance from the VR camera. Furthermore, the occlusion culling removes the rendering computation for the nodes occluded from other objects and hidden from the VR camera.

Starting from $0$, each LOD is associated with a polygonal complexity of the sphere. In particular, the LOD $0$ allows the spheres visualization in high poly with simple labels associated that represent the node's description. The latter rotates towards the VR camera so that they are always visible to the user. Labels in LOD $1$ are hidden because we assumed they were not visible beyond a significant distance from the user, and the spheres are less complex in terms of polygons. In LOD $2$, we further reduce the complexity of the sphere because there is a great distance from the user. We determine the LOD activation distance considering the percentage of the camera engaged from nodes. To obtain better performance in terms of graphics quality and rendering time, we activated the LOD $0$ when the percentage was between $100\%$ and $15\%$, the LOD $1$ when the percentage was between $15\%$ and $3\%$, and the LOD $2$ when the percentage was between $3\%$ and $1\%$. To further reduce the computational effort, we firstly hide the edges when the nodes are selected and moved in the scene by the user and then we recompute their lengths and positions when the nodes were released in the new positions. Thereby, we displayed again them, decreasing the positions computation effort which was performed only one time, and the rendering time.
After generating the graph by importing the \textit{.gexf} file, the user can select the display mode and subsequently the interaction mode (see Section~\ref{sec:back}
). The VR camera movement is allowed freely in all directions, and the movement speed can be increased or decreased. The viewfinder can be activated and moved on the target nodes enabling the ray casting using the pinch gesture or the trigger button of the HTC controller or the trackball, based on the visualization experiment the user was involved in. The viewfinder is also activated to deselect the nodes using the methods discussed in Section~\ref{sec:interaction}. It is possible to deselect all nodes by pointing the viewfinder and enabling the ray casting toward an empty scene area.

\begin{figure}[ht!]
    \centering
    \begin{subfigure}{0.475\textwidth}
    \includegraphics[width=1.0\textwidth]{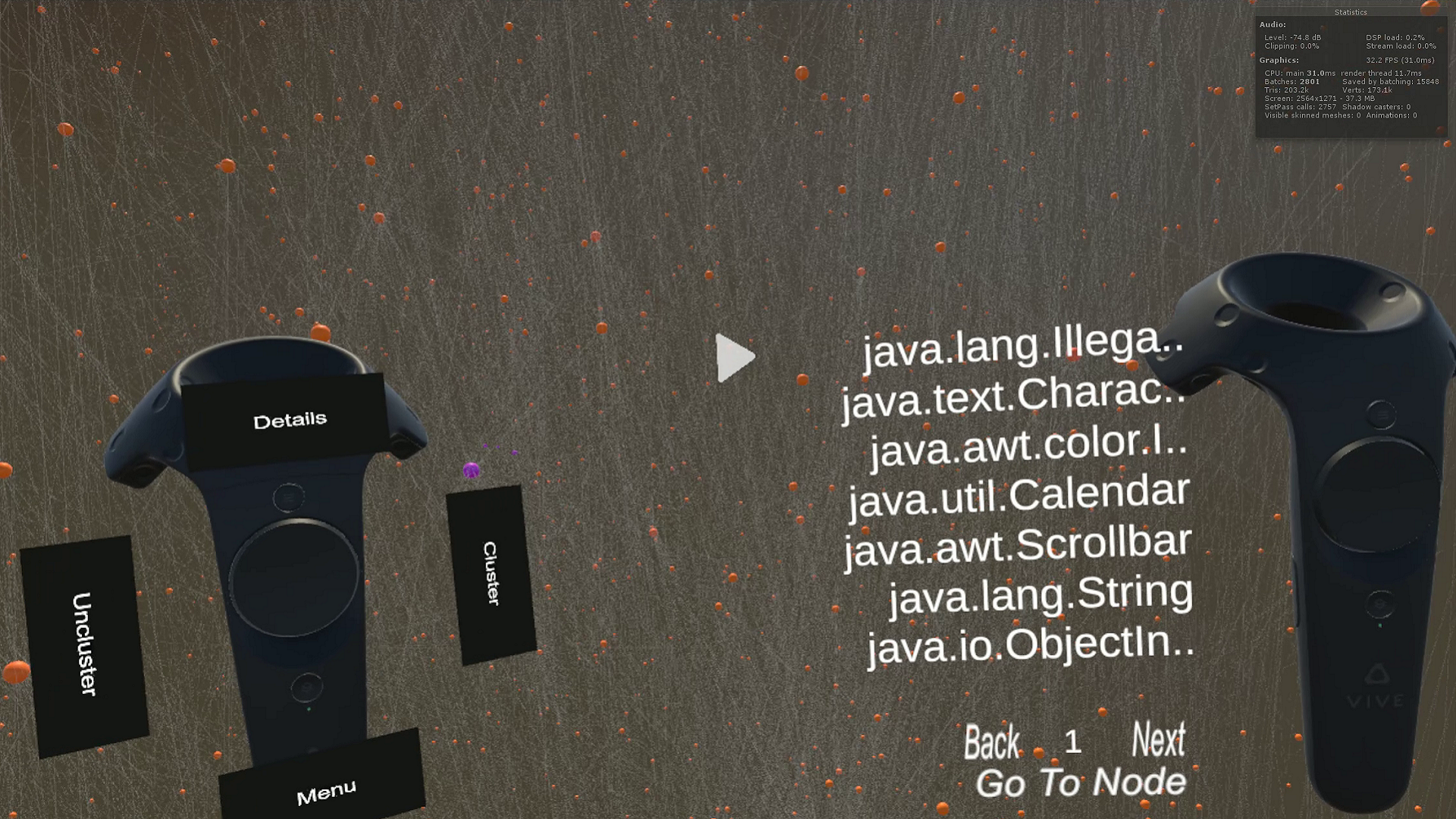}
    \caption{}
    \label{fig:menu_htc}
    \end{subfigure}
    \hfill
    \begin{subfigure}{0.475\textwidth}
    \includegraphics[width=1.0\textwidth]{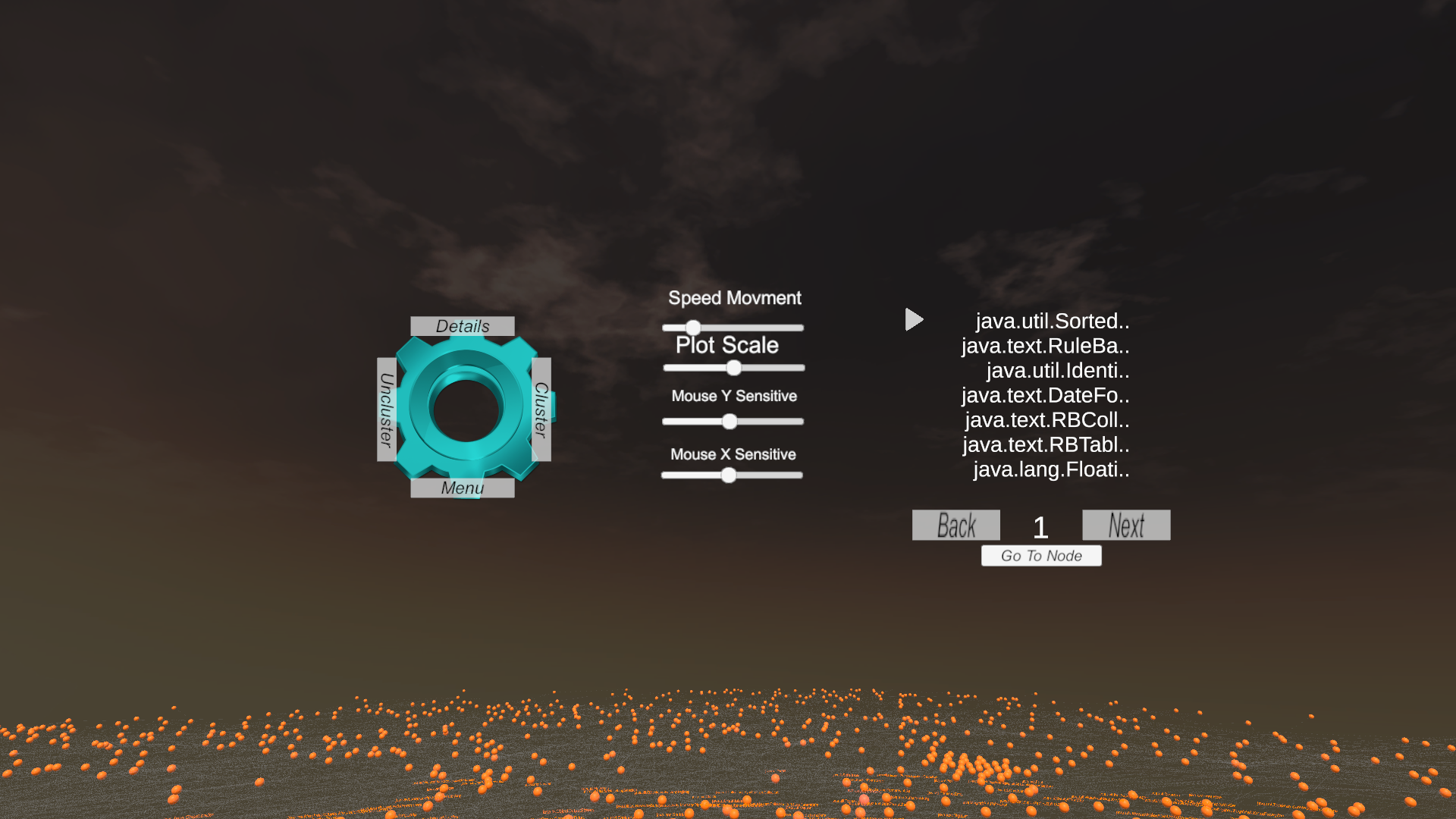}
    \caption{}
    \label{fig:menu_track}
    \end{subfigure}
    \caption{The application menu in HMD and IVE visualization modalities. Both allow the user to create a \textit{Cluster} perform the \textit{Uncluster}, see \textit{Details}, and close the \textit{Menu}. The list of nodes can be scrolled by the user, and he can teleport in front of the selected node through the \textit{Go To Node} button. In the IVE modality, the application menu shows some settings useful to change the locomotion speed, scale the graph, and change the sensibility of the trackball ball.}
    \label{fig:menus}
\end{figure}

As shown in Figure~\ref{fig:menus}, we implemented an application menu to allow settings and features. This menu allows the user to view the list of selected nodes, chose a node, and teleport himself in front of it using the \textit{Go To Node} feature. When the application menu is activated, four buttons are displayed on the HTC locomotion controller or the IVE, as shown in Figure~\ref{fig:menus}. It is possible to gather the selected nodes in a cluster node using the button \textit{Cluster}. The label related to the cluster node contains a progressive number as text. It is possible to create clusters of clusters or clusters with clusters and nodes. Likewise through the \textit{Uncluster} button is possible to split the nodes of the selected cluster. Furthermore, the button \textit{Menu} allows the user to close the menu, and the button \textit{Details} enable the user to view the selected node label or the list of nodes contained in the cluster. Finally, as shown in Figure~\ref{fig:menu_track}, the application menu in IVE mode allows changing some settings such as the movement speed, the scale, and the ball (or mouse pointer) sensibility.

\section{Evaluation Study}\label{sec:evaluationstudy}
Our evaluation study is based on the human-computer interaction methodology~\cite{lazar_research_2010}, which is applied in several contexts~\cite{Leon:2012,Blake:2014,Al-Musawi2016,ZaccagninoIV2015,ZaccagninoIV2016}. Considering the configurations shown in Fig.~\ref{fig:configurations}, we try to answer the following research questions:
\begin{figure}[h!t]
	\centering
	\includegraphics[width=1\textwidth]{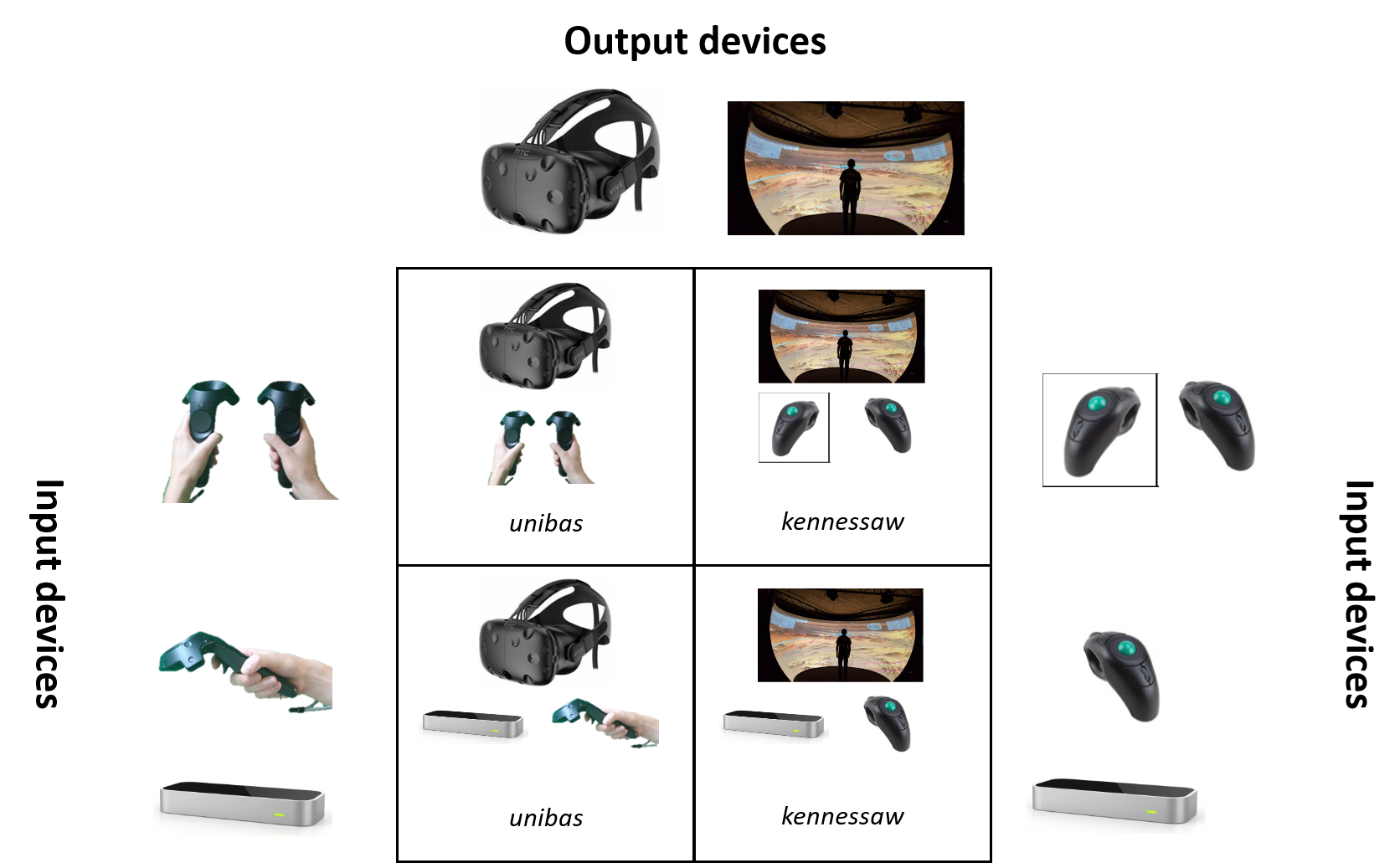}
	\caption{Two sets of configurations were tested by the two samples of enrolled participants: \textit{unibas}, from an Italian university campus and \textit{kennesaw}, from a US university campus.}
	\label{fig:configurations}
\end{figure}

\begin{itemize}
\item What differences exist, when navigating structured data in a 3D environment, between Immersive Visualization Environment-based visualization systems and HMD-based visualization systems in terms of the degree of immersion perceived by users?   
\item Which, among the four different interaction modalities experimented, is the most effective and usable?
\item Which is the reaction of users concerning the experimented interaction modalities, in terms of usefulness, easiness, and behavioral intention to use?
\end{itemize}

\subsection{Procedure} \label{subsec:procedure}
To find answers to the above questions and analyze the use of different configurations in the context of 3D data visualization, we conducted an evaluation study involving two samples of 30 participants among students from both the Italy and the US University campuses.

The Italian study was conducted in the Parallel and Computer Graphics Lab research at the University of Basilicata, Italy. 
We tested here two configurations involving the use of the HTC Vive as the output device, the use of two controllers ($HTCVive\_2Controllers$) and one controller and a Leap Motion as input devices ($HTCVive\_Controller\_LeapMotion$). We refer to this set of experiments as \textit{unibas} (see Fig.~\ref{fig:configurations}). A workstation was used, equipped with an i7, 3.60GHz, an NVidia Titan Xp GPU, and 16 GB of main memory.

The US study was conducted in the Visualization \& Simulation Research Center Cluster laboratory at the Kennesaw State University, Georgia, US. Here were tested two configurations involving the use of an IVE as output device and the use of two controllers ($IVE\_2Trackballs$) and of one controller and a Leap Motion as input devices ($IVE\_Trackball\_LeapMotion$). We refer to this set of experiments as \textit{kennesaw} (see Fig.~\ref{fig:configurations}). 

The graph consists of $1538$ nodes and $8032$ edges, rendered in a fixed frame rate of $50$ frame per second in the HMD environment, and $71$ frame per second in IVE. However, we aimed to analyze the users' interactions, locomotion, and VR visualization of the explored graph and obtain information about which interaction, locomotion, and visualization modality were most effective and usable. 

The study consisted of three phases, namely: (a) a Preliminary Survey; (b) a Testing Phase; and (c) a Summary Survey, as defined and implemented in the same and other contexts~\cite{malandrino2014, Andreoli16, ErraMP19}. In the first phase, we asked participants to fill in a preliminary questionnaire to collect: (a) demographic information (i.e., gender, age, education level); (b) information and communications technology (ICT) expertise; (c) general attitudes towards video games; and (d) general familiarity and experience with both virtual reality and graph theory. The 20 questions (listed in Appendix~\ref{A1}) 
took various forms; some were open-ended, others required a \quotes{yes} or \quotes{no} answer, others required stating a preference from up to 10 possible choices, and finally, others were rated on a 5-point Likert scale with strongly agree/strongly disagree as verbal anchors. In the Testing Phase, users were given a 10-minute training period to become familiar with the configuration. Then they were asked to complete the following three tasks: 
\begin{itemize}
  \item T1 (Search): Given the JPLD graph, \quotes{\textit{try to find four gray nodes within a fixed amount of time, and try to read their labels}}.  When the user will approach the node, it will become green. 
    \item T2 (Exploration): Given the JPLD graph, \quotes{\textit{explore it to select as many labels as possible (mild recommendation), within a fixed amount of time}}. When the user will select a node, it will assume the magenta color.  
    \item T3 (Clustering): Given the JPLD graph, \quotes{\textit{try to find and cluster gray nodes}}. A cluster is made up of at least two nodes. When a cluster is created, the corresponding nodes will group into a red node.
\end{itemize}
The first task aimed to reveal the capacity of users to orient themselves. The second aimed to test their ability (rapidity) to move through a dense graph to have an idea of the structure of the graph. We have to emphasize that labels were not counted by users, but just selected. Conversely, the system counted the number of nodes selected by the participants in the study. Moreover, labels were not selected (and therefore counted) multiple times, since, after their selection from users, the system changes their color. In this way, users are informed that the nodes cannot be selected anymore. Finally, the third task was designed to test complex user interactions. 

In all tasks, the user was not told the maximum time allowed (5 min for the Search and Clustering tasks, and 2 min for the Exploration task). This allowed us to assess the degree of the user's involvement, without creating anxiety during task execution. At the end of each task, we asked participants to rate how easy it was to perform. Other questions addressed the responsiveness of the configuration, how natural the interaction was, and finally, whether they experienced any problems during task execution (dizziness, nausea, tiredness, movement limitations, etc.). The first three questions were rated on a 5-point Likert scale, while the fourth offered up to 9 different choices (see Appendix~\ref{A2}). 
Users were monitored during the experiment and could call for assistance if they did not understand any of the instructions. Testing was performed in an isolated environment in our research lab to avoid distractions due to the presence of other people. Users were also encouraged to provide informal feedback (such as general comments or suggestions). 

At the end of the testing phase, we asked users to fill in a questionnaire, adapted from the standard \textit{Presence} Questionnaire~\cite{Witmer:1998}, aiming to assess the perceived degree of involvement and enjoyment. From this questionnaire, we extracted questions that allow us to obtain sub-scores in terms of four factors: realism (RF), control (CF), distraction (DF), and involvement (INV).  Each question was a rating on a 7-point Likert scale, with \textit{strongly agree} and \textit{strongly disagree} as verbal anchors (see Appendix~\ref{A3}). 

Finally, in the last phase, we asked users to complete a summary questionnaire. The objective is to derive participants' summary conclusions about easiness, and usefulness, as well as their interest in the idea and their behavioral intention to use such systems in the future.  This final questionnaire asked them to state their preferences from a maximum of 15 options and included three other questions rated on a 5-point Likert scale. (see Appendix~\ref{A4}). 
Each configuration was tested once by each user on the same day, and each test lasted between 45 and 50 minutes. The full test schedule in the US required one week to be completed, while the Italian test schedule required two weeks, given the breaks due to the pandemic COVID-19  period. The questionnaires that were used are reported in Appendix~\ref{appendix}.

\subsection{Recruitment}
Participants were students at two university campuses, the first one in Italy, i.e., the University of Basilicata, and the second one in the US, i.e., the Kennesaw State University. They were recruited through word-of-mouth, advertising, and student mailing lists.
Their participation was voluntary and anonymous. Participants were informed that all the information they provided would remain confidential.

\subsection{Data Analysis}
Non-parametric tests were applied to study differences between the groups testing the defined configurations. The Shapiro--Wilk goodness-of-fit test was used to assess the normality of the data~\cite{Shapiro1965}.
We recall too the reader that the \textit{p}-value is used in the context of null hypothesis testing to quantify the statistical significance and that the smaller the \textit{p}-value, the larger that significance.  
%The internal consistency reliability among the multi-item scales was examined with Cronbach's alpha~\cite{Cronbach51}. 
Finally, questionnaire responses were analyzed using SPSS version 20.\footnote{\url{http://www-01.ibm.com/software/analytics/spss/}}.

\section{Results}\label{sec:results}

In this section, we discuss the results of each of the three phases of our evaluation study.

\subsection{Preliminary Survey Results}
As shown in Table~\ref{table:Demographics}, we recruited 60 participants with bachelor's  (60\%), master's (20\%), and Ph.D.  students (7\%) from the Computer Science and the Coles College of Business departments at the University of Salerno and Kennesaw State University,  respectively. The majority were male (65\%) with the age of participants mostly in the range of 20-26 years (68.3\%).
The majority of respondents (85\%) said that they spend less than 7 hours per week playing video games, and more than half (70\%) considered themselves to be \quotes{competent/expert} in ICT.

\begin{table}[h!t]
\caption{Participant Demographics}
\centering
\fontsize{2.5mm}{2.5mm}
\selectfont{
\renewcommand{\arraystretch}{1.3}
\setlength\tabcolsep{3.0pt}
\begin{tabular}{l  c  c}
	\toprule
              &\textbf{Number}&\textbf{Percentage}\\ \textbf{Total Participants} & 60 &  \\
\midrule
 \textbf{\emph{Gender}} &  &   \\
  \hspace{0.3cm} Male    &39&  82\%\\
  \hspace{0.3cm} Female  &21&  18\%\\
  \hline
  \textbf{\emph{Age}} &&\\
  \hspace{0.3cm} 20--23 years &20& 33\% \\
  \hspace{0.3cm} 24--26 years &21& 35\%\\
  \hspace{0.3cm} 26+ years   &19 & 32\%\\
  \hline
  \textbf{\emph{Education Level Attained}} &&\\
  \hspace{0.3cm} Bachelor's, Master's, PhD             &52&87\%\\
  \hspace{0.3cm} High School                           &8& 13\%\\
  \hline
\textbf{\emph{Time Playing Video Games per Week}} &&\\
   \hspace{0.3cm} 0--7 hours  &51 &85\%\\
   \hspace{0.3cm} 8--21 hours &8 & 13\%\\
   \hspace{0.3cm} 21+ hours & 1& 2\%\\
  \hline
  \textbf{\emph{Experience in ICT field}}&&\\
   \hspace{0.3cm} Beginner  & 18 & 30\% \\
   \hspace{0.3cm} Competent & 16 & 27\% \\
   \hspace{0.3cm} Expert    & 26 & 43\% \\
\bottomrule
\end{tabular}}
\label{table:Demographics}
\end{table}

Results of the preliminary survey show that participants were not very familiar with video games (only 28\% rated themselves as \quotes{expert} in the field, 85\% spend about an hour per week playing a video game, on average) and that they mostly used traditional input devices, namely mouse and keyboard (48.3\%), and the PlayStation console (46.7\%) to play them. Moreover, when interviewed about their familiarity with natural user interfaces, only 16.7\% of participants in the study gave Leap Motion as an answer.
To derive the background of the participants concerning the domain under analysis, we submitted specific questions about the graph theory and the immersive environments. 
    When interviewed about their familiarity with graph theory, 40\% of the participants expressed high familiarity, while \quotes{orientation} was the most rated issue when interacting with a graph (13.3\%). Finally, 31.6\% of participants stated that they were familiar with Virtual Reality, 72\% of participants affirmed knowing immersive devices, but only 39\% of participants rated high his experience with them.

Three questions from the preliminary survey questionnaire (i.e., Q1, Q2, and Q3 in the Preliminary Survey Questionnaire) were supplied as input to the \emph{k}-means clustering algorithm~\cite{DataMiningBook}. As a result, we identified three groups of participants: (i) a \emph{LowGamers} group, who do not like so many video games (55\%), (ii) a \emph{MediumGamers} group, who do like video games and spend a small proportion of their time playing them (31.7\%), and (iii) a \emph{HardGamers} group, who spend a considerable proportion of their time playing video games (13.3\%). 

In summary, the first phase allowed us to build a profile of our participants. Specifically, our sample had participants with a low interest in video gaming (only 28\% rated themselves as the expert) and immersive environments (only 7\% expressed high familiarity, while almost 50\% of participants rated themselves as inexpert in that field), with high technical skills (being Computer Science students) and low familiarity with graph theory (beginner the 60\% of our participants). 
Finally, 15\% of participants were left-handed, 8\% suffered from motion sickness, 22\% had vision deficiencies (myopia), and more than half of the participants wore glasses.

\subsection{Testing Phase Results}
\begin{figure}[h!t]
	\centering
	\includegraphics[width=9cm]{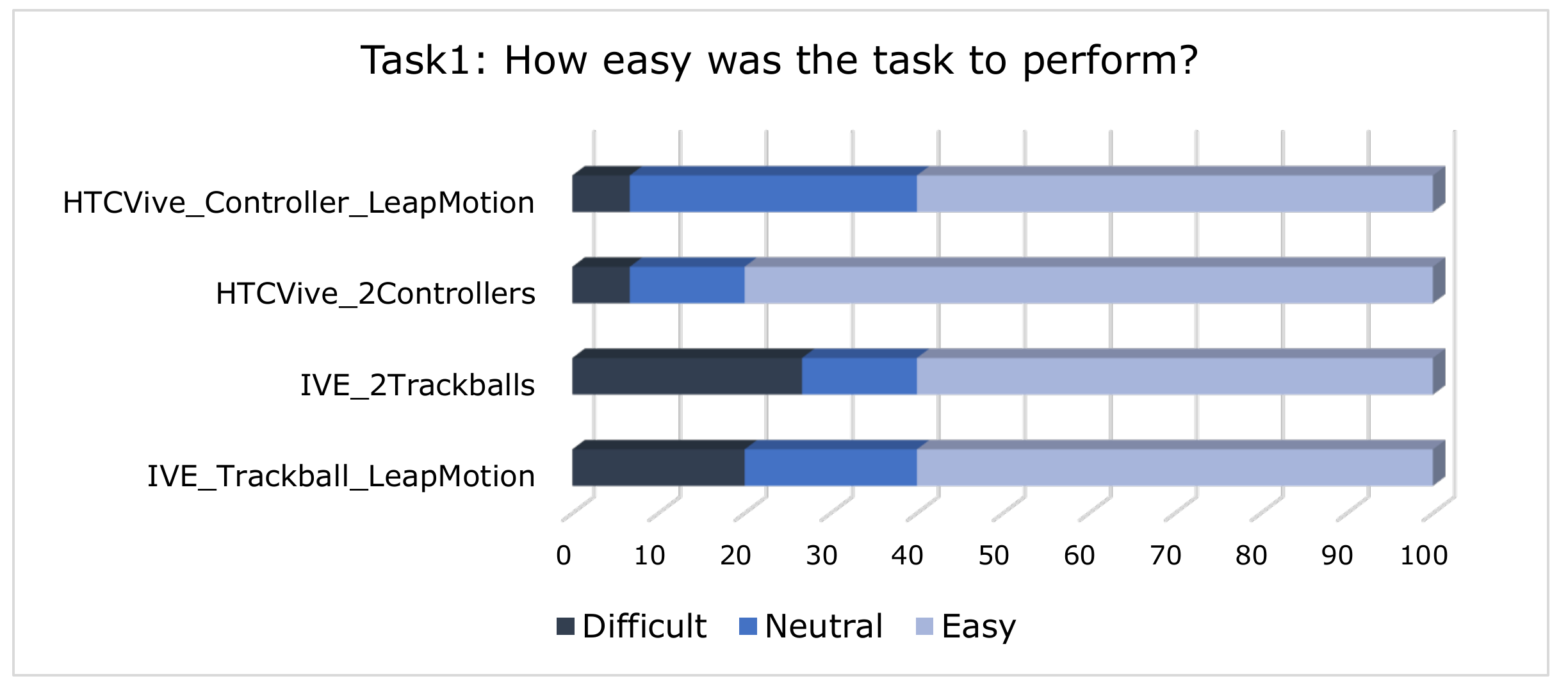}
	\caption{Comparison of tested configurations in terms of easiness of performing tasks.}
	\label{fig:easiness}
\end{figure}

\begin{figure}[h!t]
	\centering
	\includegraphics[width=9cm]{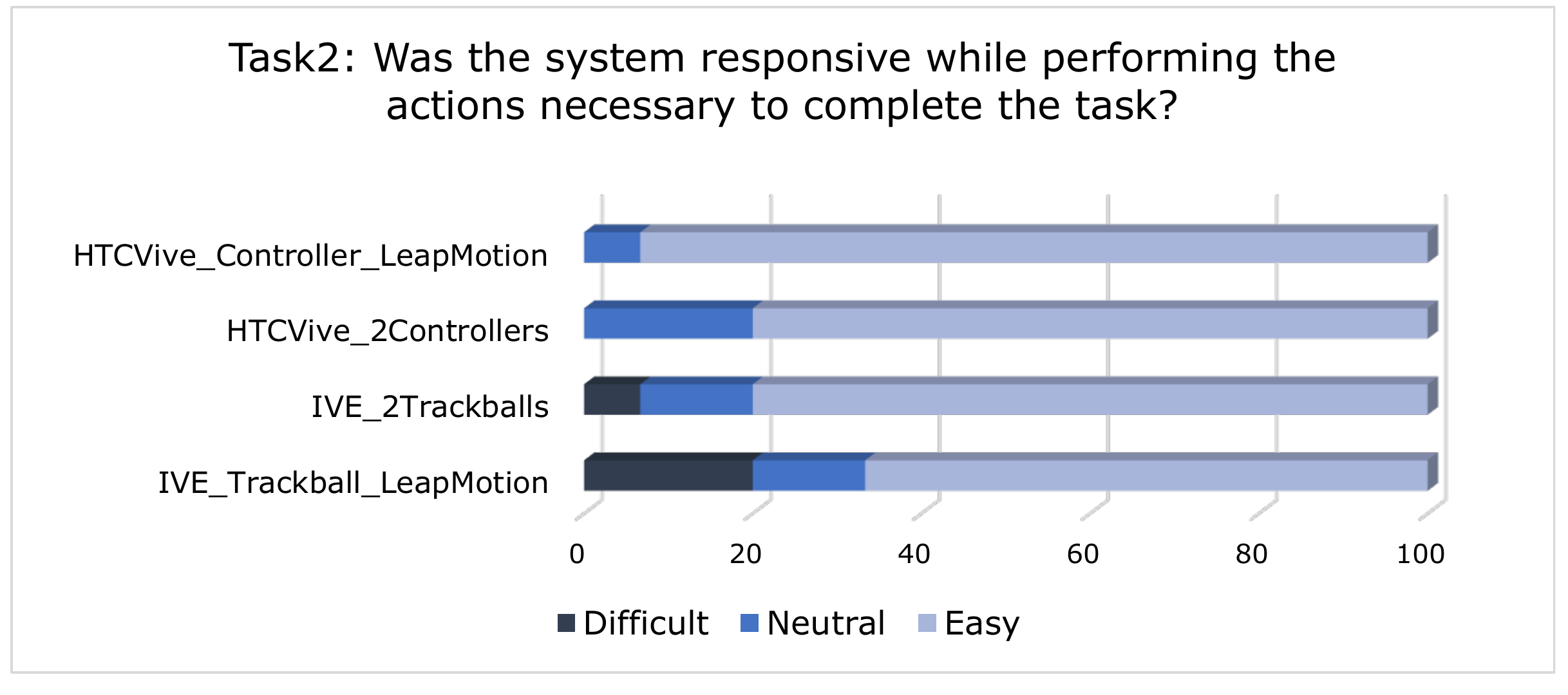}
	\caption{Comparison of tested configurations in terms of responsiveness of performing tasks.}
	\label{fig:responsiveness}
\end{figure}

\begin{figure}[h!t]
	\centering
	\includegraphics[width=9cm]{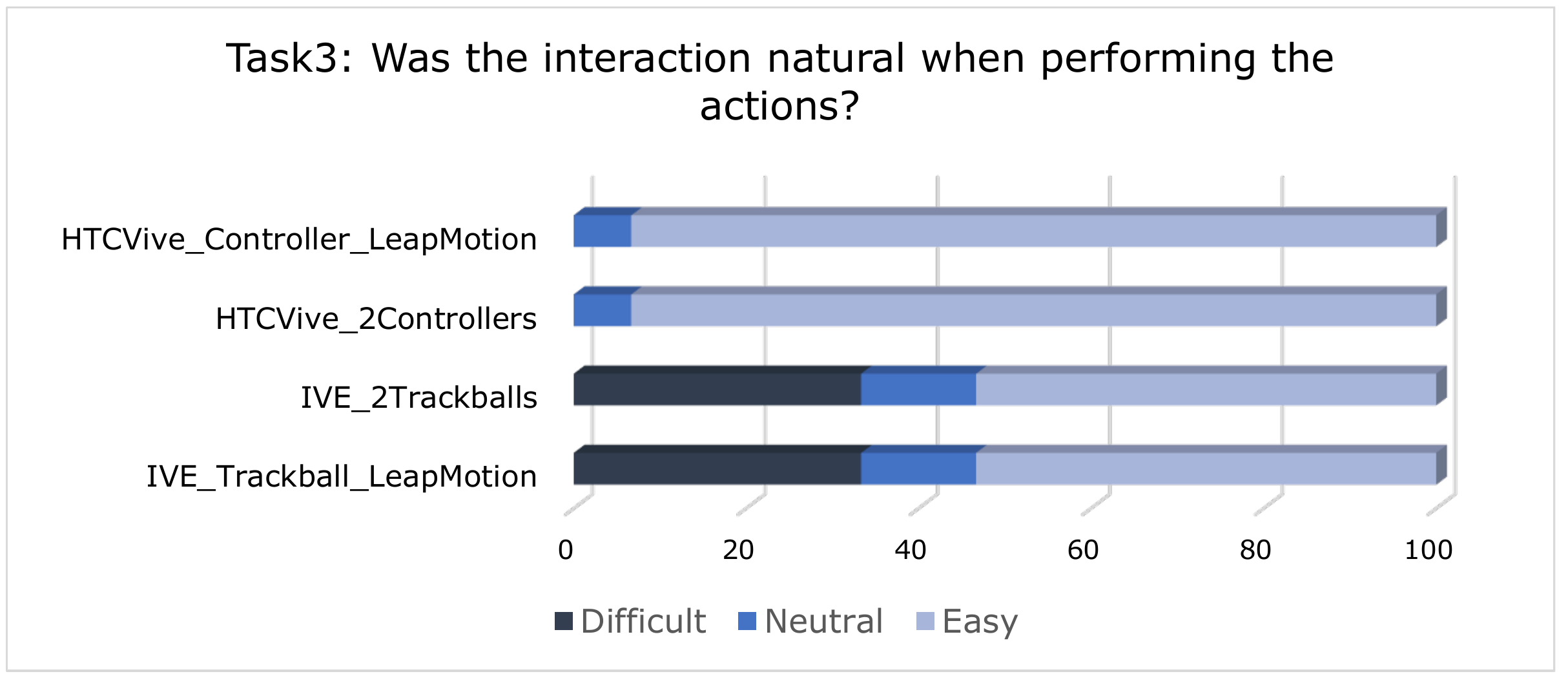}
	\caption{Comparison of tested configurations in terms of how natural were the interactions while performing tasks.}
	\label{fig:natural}
\end{figure}

The second phase involved interaction with the developed system.
Recall that participants were asked to perform three tasks and to evaluate afterwards their easiness (Fig.~\ref{fig:easiness}), the responsiveness of the overall configuration (Fig.~\ref{fig:responsiveness}), and how natural the interactions felt during the testing of the configuration  (Fig.~\ref{fig:natural}).
As we can see, all configurations were evaluated positively, even if HMD-based configurations (i.e., $HTCVive\_2Controllers$ and $HTCVive\_Controller\_LeapMotion$) obtained slightly better results across all tasks. Interestingly, no participants who tested HMD-based configurations rated negatively (\quotes{Low} value for Task 2, \quotes{Difficult} value for Task3) the responsiveness and the naturalness of the interactions. We also analyzed whether differences existed between the 4 groups (configurations) and whether these were statistically significant. We found that groups did not differ regarding the easiness, responsiveness, and naturalness of interactions tested in the three assigned tasks.

During this phase, sessions were monitored to measure the time required by participants to complete the tasks and to measure the corresponding precision/correctness.
We have to emphasize that for Task 2 we did not measure the completion time, since here we asked participants to explore as many nodes as possible in the fixed time of 2 minutes.
About the correctness, for Task 1, the correctness is given by the number of nodes that participants were able to find, while for Task 3, the precision is given by the number of clusters that participants were able to build. Task2 did not require any analysis of correctness, since here we were interested in deriving information about the rapidity of participants while exploring the graph.

\begin{figure}[h!t]
	\centering
	\includegraphics[width=9cm]{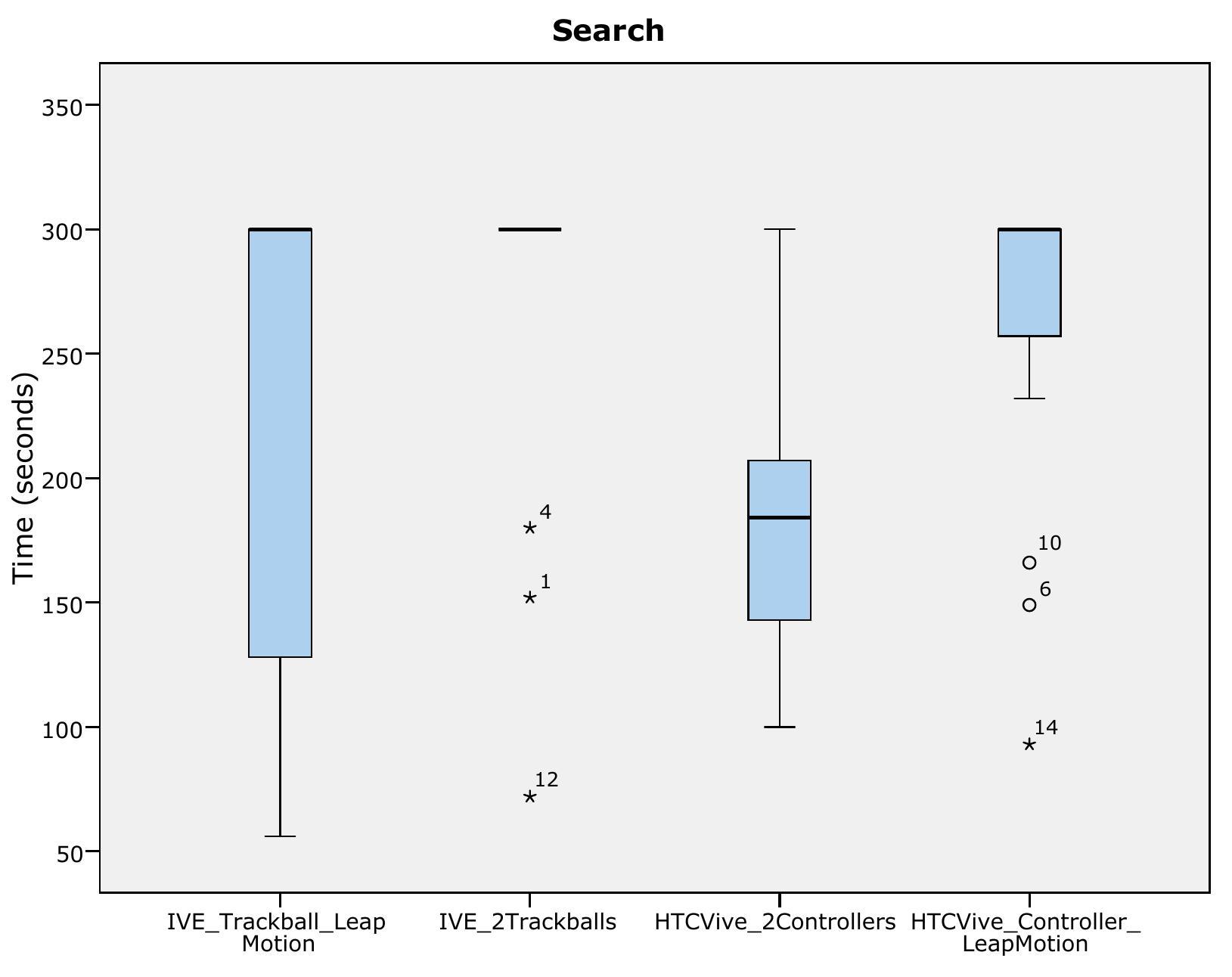}
	\caption{Comparison of tested configurations in terms of Time to complete Task1.}
	\label{fig:TimeTask1_All}
\end{figure}

As we can see in Fig.~\ref{fig:TimeTask1_All}, about Task1, participants who tested $HTCVive\_2Controllers$ spent less time completing the task.  The worse result was obtained by participants testing $IVE\_2Trackballs$, where only 3 participants out of 15 were able to complete the task without reaching the time limit of 5 minutes. This difference among configurations is also statistically significant  ($p<0.012$).
About completion time for Task 3, we observed that all participants except those who tested the $HTCVive\_2Controllers$ reached the maximum allowed time to perform the task. The HMD-based configuration employing two controllers as input devices ensured the completion of the task within the time limit (M=271.3 seconds, SD=64.6 seconds). 

When analyzing precision/correctness, we found out, concerning Task 1, that the highest precision was exhibited by participants testing $HTCVive\_2Controllers$, where indeed 87\% of them completed the task with no error.  The second-best performing configuration was, $IVE\_Trackball\_LeapMotion$, with 47\% of participants performing the task with 0 errors. 

\begin{figure}[h!t]
	\centering
	\includegraphics[width=9cm]{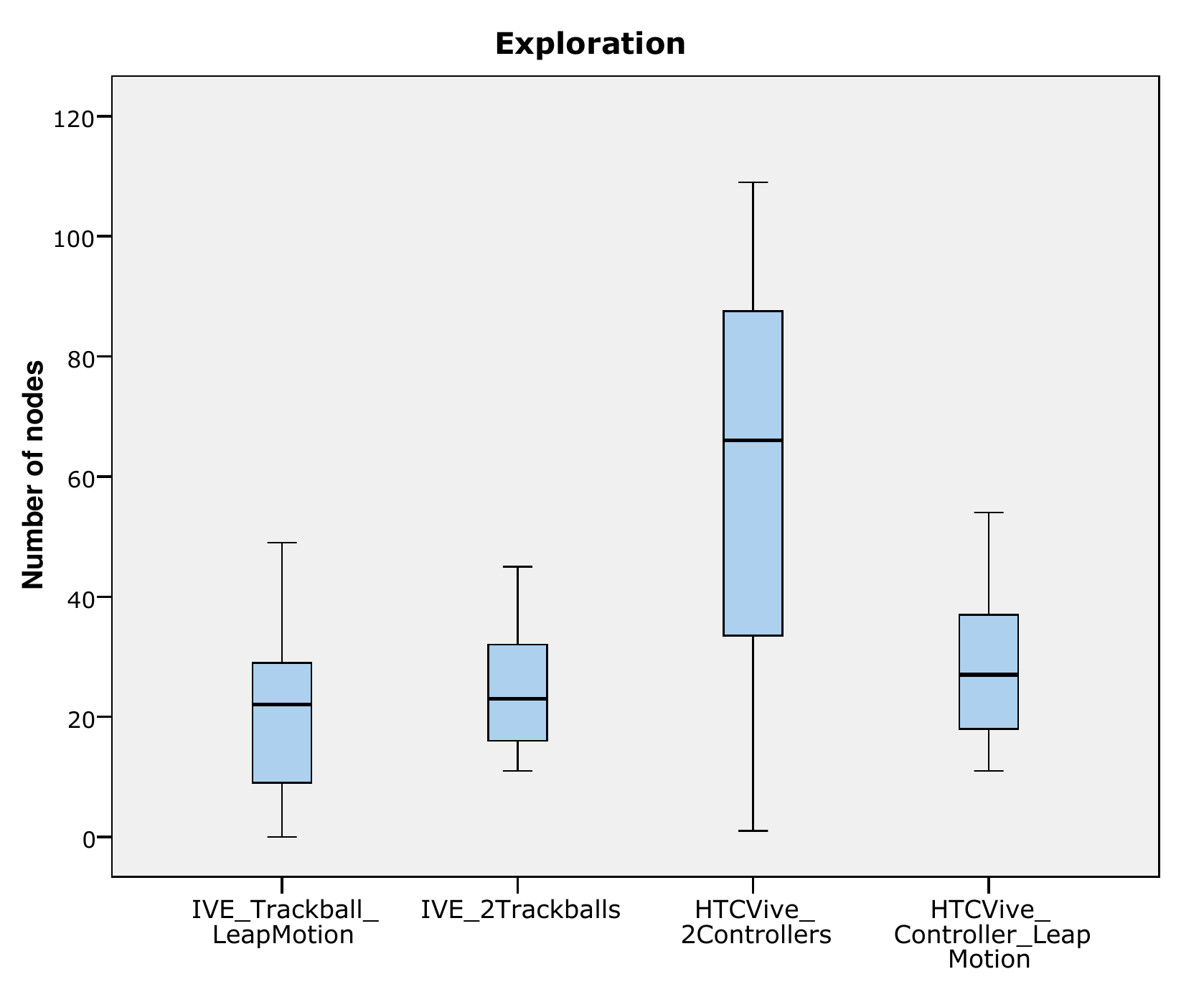}
	\caption{Comparison of tested configurations in terms of accuracy when exploring nodes in Task 2.}
	\label{fig:CorrectnessTask2}
\end{figure}

\begin{figure}[h!t]
	\centering
	\includegraphics[width=9cm]{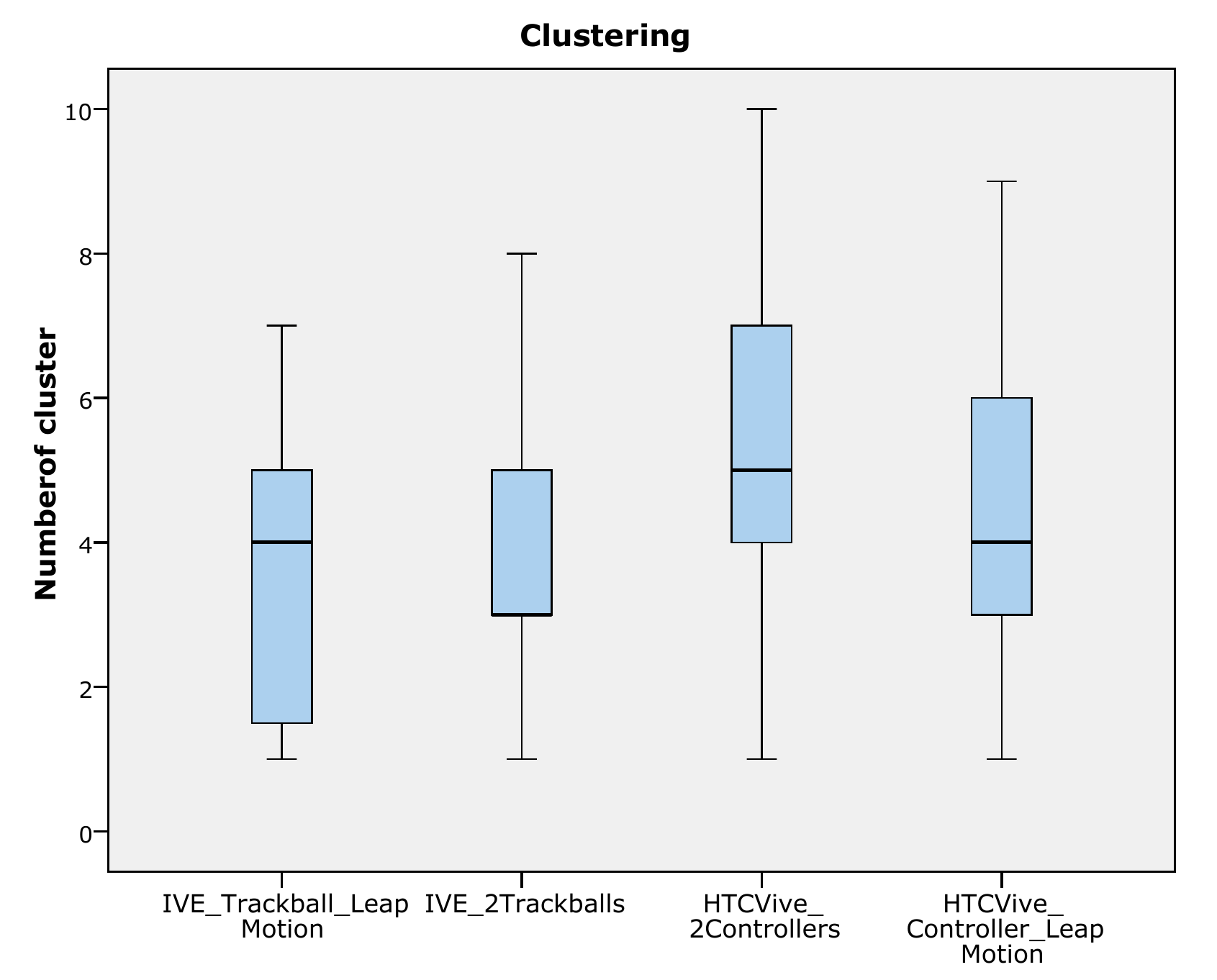}
	\caption{Comparison of tested configurations in terms of accuracy when clustering nodes in Task 3.}
	\label{fig:CorrectnessTask3}
\end{figure}

With regard to Task 2, in Fig.~\ref{fig:CorrectnessTask2} we show how participants testing  $HTCVive\_2Controllers$ were able to find the highest number of nodes. We also found a significant statistical difference among the groups (p$<$0.01).
Finally, with regard to Task 3, we observed that even here, the participants testing $HTCVive\_2Controllers$ performed better, building the highest number of clusters (see  Fig.~\ref{fig:CorrectnessTask3}).

Finally, we did not find any statistical difference among the three video gamers groups (\emph{NoGamers}, \emph{LowGamers}, and \emph{HardGamers}) with regard to both time and precision/correctness for Task 1.
Conversely, when analyzing video gamers groups with regard to the number of nodes explored in Task 2 and the number of clusters built-in Task 3 we found significant statistical differences (p$<.01$ for both). The insight here is that video gamers were more skillfully in activities requiring the selection of the nodes and their clustering (complex operations).

In summary, the analysis of the Testing phase ended with the interesting result that the configuration involving the use of the HTC Vive as an output device and 2 controllers as input devices shows the best results in terms of both performance and correctness. Additionally, this result is still valid with regard to the analysis of the subjective metrics envisioned in our study (easiness, responsiveness, and naturalness of interactions of the performed tasks), with better results obtained by participants who tested the $HTCVive\_2Controllers$ configuration.

As described in Section~\ref{subsec:procedure}, to analyze the opinions of participants in terms of immersion and their overall subjective experience, we used the standard Presence Questionnaire (PQ). It measures to what degree users feel as if they are part of the experienced environment. From this questionnaire, we extracted questions that allow obtaining sub-scores in terms of four factors: realism (RF), control factor (CF), Distraction Factor (DF), and involvement (INV). 

\begin{figure}[h!t]
	\centering
	\includegraphics[width=9cm]{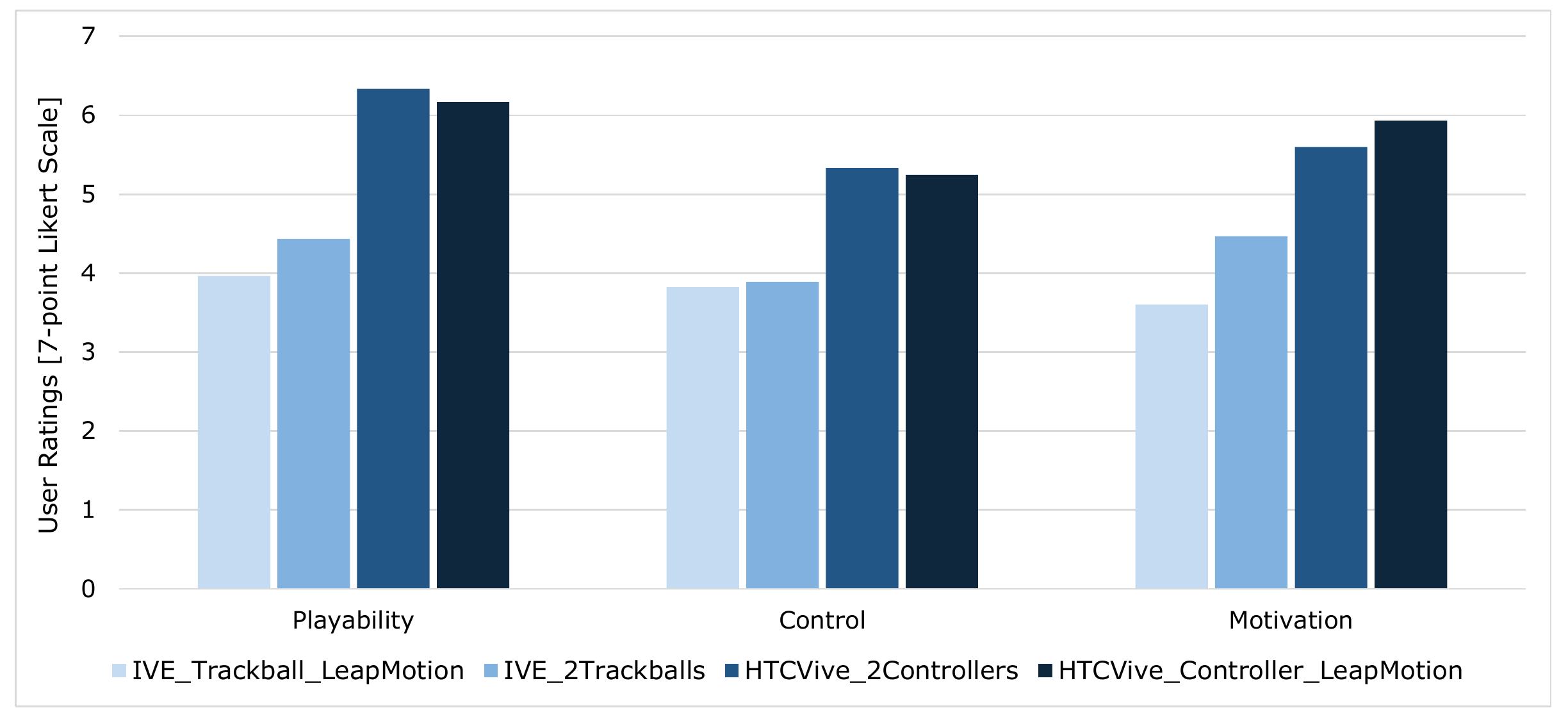}
	\caption{Playability, control, and motivation. Questions in Appendix~\ref{A3}).}
	\label{fig:PCM}
\end{figure}

First, we report in Fig.~\ref{fig:PCM} the results of users' perceptions about playability, the control they had within the immersive experience (represented by their tasks), and their motivation. Regarding playability, all participants enjoyed the experience, with the best scores obtained with participants testing the $HTCVive\_2Controllers$ (M=6.3, SD=0.7) and $HTCVive\_Controller\_LeapMotion$ (M=6.17, SD=0.79) configurations, with a significant statistical difference (p$<$.0001). 
About the \textit{Control} factor, participants rated positively the aesthetics of the virtual representation (M=5.0, SD=1.6), with slightly better results achieved with HMD-based configurations. We also found a statistical difference across the configurations (p$<$.01).  Moreover, they agreed with how easy it was to use the immersive devices, M=3.7, SD=2.2) and expressed high satisfaction in terms of the sensitivity of the controllers (\quotes{\textit{The game controller sensitivity in the virtual experience was adequate}}, M=5.1, SD=1.6). Similarly, the \textit{Motivation} factor was positively rated by our sample (M=4.9, SD=1.8).  Even here, better results were obtained by participants testing HMD-based configurations, with a significant statistical difference among the 4 groups (p$<$.01). 

As shown in Table~\ref{tab:PresenceImmersion}, the Presence questionnaire is aggregated into four factors, almost all positively rated. Participants were found out responsive to the overall environment (M=5.1, SD=0.7), and they perceived very natural interactions within the immersive experience (M=4.8, SD=1.1).  Additionally, they rated very positively their involvement (M=5.3, SD=0.5), until to the extent they lost track of the time (M=5.0, SD=0.4). 
Less positive results were obtained with questions related to the Distraction and Realism factors, with mostly disagree ratings about the questions: \quotes{\textit{To what degree did you feel confused or disoriented at the beginning of breaks or the end of the experimental session?''} (M=2.8,SD=0.2), \quotes{\textit{How distracting was the control mechanism?}}} (M=2.6, SD=0.4), and finally, 
\quotes{\textit{How much did the control devices interfere with the performance of assigned tasks or with other activities?}} (M=2.7,SD=0.3).
Similar to the previous analysis, questions were positively rated mostly by participants testing the $HTCVive\_2Controllers$ and $HTCVive\_Controller\_LeapMotion$) configurations.

\begin{table*}[h!t] 
	\centering        
	\fontsize{2.5mm}{2.5mm}\selectfont{
		\renewcommand{\arraystretch}{1.3}
		\setlength\tabcolsep{10pt}
		\caption{Presence and Immersion. 7-Point Likert scale.\label{tab:PresenceImmersion}}{
			\begin{tabular}{cp{4cm}|c|c|c|c|c}
				\toprule
				ID &Question  &IVE\_                 &IVE\_        & HTCVive\_           &HTCVive\_ &Factor \\
			       &          &Trackball\_ &2Trackballs  &  2Controllers       & Controller\_ \\
			       &          &LeapMotion &            &                     & LeapMotion\\
				P1&How responsive was the environment to actions that you initiated (or performed)? &4.33 &  4.80    & 5.80 & 5.60 & CF\\
				P2&How natural did your interactions with the environment seem?   &3.67    &   3.93    & 5.87  & 5.60  &CF \\
				P3&How involved were you in the virtual environment experience?   &4.80    &   5.07    & 5.87  & 5.67  &INV\\
				P4&Were you involved in the experimental task to the extent that you lost track of time?  &4.40   &   5.13    & 4.87 &5.47&INV  \\
				P5&How proficient in moving and interacting with the virtual environment did you feel at the end of the experience? &4.40  &   4.33   & 5.53  &5.93&CF \\
				P6&How aware were you of events occurring in the real world around you?  &3.53&    2.67   & 4.20 &4.00 &DF\\
				P7&How aware were you of your display and control devices?&3.60&1.73&5.27&4.73&DF\\
				P8&To what degree did you feel confused or disoriented at the beginning of breaks or at the end of the experimental session? &2.80&2.53&2.73&3.07&RF\\
				P9&How distracting was the control mechanism? &2.73&2.00&2.93&2.87&DF\\
				P10&How much did the control devices interfere with the performance of assigned tasks or with other activities?&2.67&2.27&3.00&3.00&DF, CF\\
				\bottomrule
			\end{tabular}}}
		\end{table*}%
		
\subsection{Summary Survey Results}\label{subsec:summarysurveyresults}
In this section, we report the results of the questions posed in the questionnaire submitted in the last phase of our evaluation study (see Table~\ref{tab:summaryresults}).
Generally, all participants rated as positive the usefulness (M=3.0) and the ease of use (M=3.2) of the tested configurations.
At the question: \quotes{\textit{Do you think the proposed system was interesting?}}, 33\% of participants testing the $IVE\_2Trackballs$ configuration expressed their disagreement. Conversely, participants testing the $HTCVive\_2Controllers$ and $HTCVive\_Controller\_LeapMotion$ configurations highly positively rated all questions. Indeed, we found a significant statistical difference among the groups concerning all three questions (last column in Table~\ref{tab:summaryresults}). 
Finally, both future own personal use and the propensity to recommend to others the experience were positively rated.   
Participants testing HMD-based configurations had a higher propensity, and indeed we found a significant statistical difference for both questions (see Table~\ref{tab:summaryresults}). 

\begin{table*}[h!t]
  \centering
   \caption{Summary survey results. 5-point Likert scale (SQ1, SQ2, SQ3), 7-point Likert scale (SQ4, SQ5). Mean (standard deviation) results.}
  \label{tab:summaryresults}
 \renewcommand{\arraystretch}{1.3}
 \setlength\tabcolsep{5.0pt}
	{
    \begin{tabular}{lp{5cm}cccccc}
    \toprule
    ID & Question     &IVE\_                 &IVE\_        & HTCVive\_           &HTCVive\_   & P value\\
	 &		       &Trackball\_              &2Trackballs  & 2Controllers        &Controller\_\\
	 &             &LeapMotion               &             &                     &LeapMotion\\
    \midrule
    SQ1&Do you think the proposed system was useful to use?& 3.0 (1.4)   & 2.9 (1.6)& 4.3 (0.6) & 4.0 (0.8) & $<$ .01 \\ 
    SQ2& Do you think the proposed system was interesting?  & 3.4 (1.5)   & 2.9 (1.7)& 4.7 (0.5) & 4.7 (0.5) & $<$ .001  \\
    SQ3&Do you think the proposed system was easy to use?  & 3.2 (1.5)   & 3.3 (1.5)& 4.4 (0.6) & 4.4 (0.6) & $<$ .05 \\
    \hline
    SQ4 & Would you take into consideration the possibility of continuing to use the system in the near future? & 3.3 (1.9) & 3.7 (2.6) & 3.7 (1.6) & 5.3 (1.6) & $<$ .05\\
    SQ5 & I will strongly recommend others to use the system    & 3.4 (1.8)& 3.9 (2.5)&6.2 (0.9) & 6.2 (1.3) &$<.001$\\

    \bottomrule
    \end{tabular}}%
	
\end{table*}%

\section{Conclusions and Future Works}\label{sec:conc}
We investigated the effectiveness of spherical-based and HMD-based systems for VR immersion focusing on different methods to explore and interact with the 3D graphs. In particular, we proposed a user study to compare the use of two HTC Vive 6-DOF controllers and a multimodal approach based on a single 6-DOF controller and the Leap Motion device for the user locomotion and interaction in a VR system based on HMD. We try to highlight the differences among these approaches in terms of the level of immersion perceived by the users, the usefulness, easiness, and behavioral intention to use our system.
This study was conducted at the University of Basilicata (Italy) by involving $30$ participants and asked them to locomote and interact with the 3D graph using the previously mentioned approaches. A similar study was conducted in parallel at the Kennesaw State University (USA) but using the IVE spherical-based system as a VR environment. In this parallel study conducted with other $30$ participants, a comparison between two trackballs and one trackball and a Leap Motion device was performed. Participants were students of both universities, recruited through advertising, word-of-mouth, and student mailing lists. As a case study, we proposed the VR 3D graph visualization application and provided details on the hand-based, controller-based, and trackball-based methods to interact with nodes, visualize their details, and locomote in the 3D scene using the controllers, hands, or trackball. Our proposed study is based on three phases: a preliminary survey in which we asked the participant to provide demographic information, ITC expertise, general video games, and attitude toward VR and graph theory; in the testing phase we provided a 10-minute training period to allow the user the configuration familiarization. Then we asked the participants to complete three tasks at different times as reported in Section~\ref{subsec:procedure} and at the end of each task, we asked participants several questions about the easiness the responsiveness of the configuration, how natural the interaction was, \etc At the end of the testing phase, we asked the participants to fill a \textit{Presence}-like questionnaire; Finally, we asked the participants to complete a summary questionnaire which aim was to investigate the easiness, the usefulness, and their possible intention to use our system in future.
The results of the preliminary investigation suggest that most of the participants are not particularly fond of video games, allowing us to define a profile of our participants. The results of the testing phase suggest that $HTCVive_2Controllers$ achieve the best performance and correctness, by matching the analysis of subjective metrics such as easiness, responsiveness, and naturalness of the performed tasks. Interesting is also the second-best configuration achieved by the$ IVE_Trackball_LeapMotion$ approach, and regardless the Task 1, the participants perform $0$ errors. About the PQ, the playability, control, and motivation in HMD configuration. Finally, the summary survey results report positive feedback about the usefulness and ease of use in all tested configurations. Furthermore, the participants appear to be inclined to use our system in the future and to positively recommend the same experience to others.
To provide a well-designed case study with a good level of rendering performance and graph complexity, we provided also simple computer graphics techniques to render as many as possible nodes and edges. In future work, we will investigate the use of deep learning-based systems to replace the Leap Motion controller with a simple RGB camera~\cite{Gruosso2020, Vakunov2020} and extend these approaches by introducing a new gesture recognition layer~\cite{caputo2021}. We will try to define the best type of hand gestures for interaction and locomotion within environments based on eXtended reality (XR) and we will try to propose a similar work with these new technologies trying to investigate the effectiveness of the use of the new paradigm of touchless interaction.

\section*{Acknowledgments}
This effort was supported by an equipment grant from the U.S. Department of Defense/U.S. Army Research Office (ARO), a National Security Agency (NSA) grant, and technical support by Immersive Display Systems Inc. The content of this work does not reflect the position or policy of the ARO, NSA, or IDS and no official endorsement should be inferred. We sincerely thank Professor James \quotes{Wes} Rhea for his professional support.

%Bibliography
\bibliographystyle{unsrt}  
\bibliography{references}

\appendix

\section{Appendix}\label{appendix}

\subsection{Preliminary survey questionnaire}\label{A1}
\begin{itemize}
\item Q1: Do you like videogames? 
\subitem Strongly disagree \radiobutton\radiobutton\radiobutton\radiobutton\radiobutton~Strongly agree

\item Q2: How do you rate your experience with video games? 
\subitem Inexpert \radiobutton\radiobutton\radiobutton\radiobutton\radiobutton~Expert

\item Q3: How many hours per week do you spend playing video games?
\subitem \radiobutton Less than on hour \radiobutton Between one and seven hours \subitem \radiobutton Between eight and fourteen hours \radiobutton Between fifteen and twenty hours \radiobutton More then twenty one hours

\item Q4: Which type of video games do you play with?
\subitem $\square$~First-Person-Shooter \xspace$\square$~Adventure \xspace$\square$~Fighting \xspace$\square$~Role-Playing Games \xspace$\square$~Strategy/Tactics \xspace$\square$~Sports
\subitem $\square$~Dance/Rhythm \xspace$\square$~Survival Horror \xspace$\square$~None (I don't play any video game) \subitem $\square$~Other (Add here which one)

\item Q5: Which type of device do you use when playing video games?
\subitem $\square$~Mouse/Keyboard \xspace$\square$~Kinect \xspace$\square$~Joypad \xspace$\square$~PlayStation \xspace$\square$~None (I don't play any video game)  \subitem $\square$~Other (Add here which one)

\item Q6: Are you left-handed?
\subitem Yes \radiobutton No \radiobutton

\item Q7: Do you suffer from kinetosis or motion sickness?
\subitem \textit{Same answer options as} Q6

\item Q8: In which of the following conditions do you feel discomfort?
\subitem $\square$~Car \xspace$\square$~Car (Rear seat/passenger seat) \xspace$\square$~Boat \xspace$\square$~Airplane \xspace$\square$~Other (Add here which one)

\item Q9: Do you have vision deficiencies?
\subitem \textit{Same answer options as} Q6

\item Q10: If you answered \quotes{Yes} to the previous question, Which type of deficiency do you have?
\subitem Open Question

\item Q11: Do you wear glasses?
\subitem \textit{Same answer options as} Q6

\item Q12: How do you rate your IT experience?
\subitem \textit{Same answer options as} Q2

\item Q13: Have you heard of the term \quotes{graph}?
\subitem \textit{Same answer options as} Q6

\item Q14: How do you rate your experience with graph theory?
\subitem \textit{Same answer options as} Q2

\item Q15: With which these natural user interfaces do you have more confidence?
\subitem \radiobutton Kinect \radiobutton LeapMotion \radiobutton PlayStation Move \radiobutton None (No familiarity with natural user interfaces) 
\subitem \radiobutton Other (Add here which one)

\item Q16: Do you have familiarity with virtual reality?
\subitem \textit{Same answer options as} Q1

\item Q17: Have you ever heard of immersive devices (i.e. HTC Vive, Oculus Rift, IVE)?
\subitem \textit{Same answer options as} Q6

\item Q18: If you answered \quotes{Yes} to the previous question, which is your experience with immersive devices?
\subitem \textit{Same answer options as} Q2

\item Q19: Have you ever needed to interact and visualize data on a graph?
\subitem \textit{Same answer options as} Q6

\item Q20: Which type of problems did you encounter? (multiple choices are allowed)
\subitem $\square$~Interaction problems (in terms of personal difficulties) \xspace$\square$~Comprehension \xspace$\square$~Information overloading \subitem $\square$~Few useful information \xspace$\square$~Orientation \xspace$\square$~None \xspace$\square$~Other (Add here which one)

\end{itemize}

\subsection{Task activities questionnaire}\label{A2}
\textit{Task One. Given the JPLD graph, \quotes{\textit{try to find four gray nodes within a fixed amount of time, and try to read their labels.}}}

\begin{itemize}
    \item T1\_a: How easy was the task to perform?
    \subitem Very difficult \radiobutton\radiobutton\radiobutton\radiobutton\radiobutton~Very easy
    
    \item T1\_b: Was the system responsive while performing the actions necessary to complete the task?
    \subitem Strongly disagree \radiobutton\radiobutton\radiobutton\radiobutton\radiobutton~Strongly agree

    \item T1\_c: Was the interaction natural when performing the actions?
    \subitem Strongly disagree \radiobutton\radiobutton\radiobutton\radiobutton\radiobutton~Strongly agree
    
    \item T1\_d: Have you experienced any of the following contraindications? 
    \subitem $\square$~High sensibility \xspace$\square$~Lack of the field of view of the device \xspace$\square$~Movements limitations \subitem $\square$~Motor difficulties \xspace $\square$~Dizzines \xspace$\square$~Nausea \xspace$\square$~Tiredness \xspace$\square$~None \xspace$\square$~Other\\
\end{itemize}
\textit{Task Two. Given the JPLD graph, \quotes{\textit{explore it to select as many labels as possible (mild recommendation), within a fixed amount of time}}. When the user will select a node, it will assume the magenta color.}
\begin{itemize}
    \item T2\_a: How easy was the task to perform?
    \subitem \textit{Same answer options as} T1\_a
    
    \item T2\_b: Was the system responsive while performing the actions necessary to complete the task?
    \subitem \textit{Same answer options as} T1\_b

    \item T2\_c: Was the interaction natural when performing the actions?
    \subitem \textit{Same answer options as} T1\_c
    
    \item T2\_d: Have you experienced any of the following contraindications? 
    \subitem \textit{Same answer options as} T1\_d\\
\end{itemize}
\textit{Task Three. Given the JPLD graph, \quotes{\textit{try to find and cluster gray nodes}}. A cluster is made up of at least two nodes. When a cluster is created, the corresponding nodes will group into a red node.}
\begin{itemize}
    \item T3\_a: How easy was the task to perform?
    \subitem \textit{Same answer options as} T1\_a
    
    \item T3\_b: Was the system responsive while performing the actions necessary to complete the task?
    \subitem \textit{Same answer options as} T1\_b

    \item T3\_c: Was the interaction natural when performing the actions?
    \subitem \textit{Same answer options as} T1\_c
    
    \item T3\_d: Have you experienced any of the following contraindications? 
    \subitem \textit{Same answer options as} T1\_d

\end{itemize}

\subsection{Presence Questionnaire}\label{A3}
Based on: Bob G. Witmer and Michael J. Singer. 1998. Measuring Presence in Virtual Environments: A Presence Questionnaire. Presence: Teleoper. Virtual
Environ. 7, 3 (June 1998), 225–240

\begin{enumerate}
    \item I enjoyed the virtual experience
    \subitem Strongly disagree \radiobutton\radiobutton\radiobutton\radiobutton\radiobutton\radiobutton\radiobutton~Strongly agree
    
    \item Exploring and playing the VisGraph3D system made me happy
    \subitem \textit{Same answer options as} 1
    
    \item I found the appearance (aesthetics) of the VisGraph3D  system to be very impressive
    \subitem \textit{Same answer options as} 1
    
    \item I knew how to use the immersive devices with the VisGraph3D system
    \subitem \textit{Same answer options as} 1
    
    \item How responsive was the environment to actions that you initiated (or performed)?
    \subitem \textit{Same answer options as} 1
    
    \item How natural did your interactions with the environment seem?
    \subitem \textit{Same answer options as} 1
    
    \item The game controller sensitivity in the virtual experience was adequate
    \subitem \textit{Same answer options as} 1
        
    \item I found the exploration with the immersive devices motivating
    \subitem \textit{Same answer options as} 1
    
    \item How involved were you in the virtual environment experience?
    \subitem \textit{Same answer options as} 1
        
    \item Were you involved in the experimental task to the extent that you lost track of time?
    \subitem \textit{Same answer options as} 1
    
    \item How proficient in moving and interacting with the virtual environment did you feel at the end of the experience?
    \subitem \textit{Same answer options as} 1
        
    \item How aware were you of events occurring in the real world around you?
    \subitem \textit{Same answer options as} 1
    
    \item How aware were you of your display and control devices?
    \subitem \textit{Same answer options as} 1
        
    \item To what degree did you feel confused or disoriented at the beginning of breaks or at the end of the experimental session?
    \subitem \textit{Same answer options as} 1
    
    \item How distracting was the control mechanism?
    \subitem \textit{Same answer options as} 1
        
    \item How much did the control devices interfere with the performance of assigned tasks or with other activities?
    \subitem \textit{Same answer options as} 1
    
    \item I will use the system on a regular basis in the future
    \subitem \textit{Same answer options as} 1
        
    \item I will strongly recommend others to use the system
    \subitem \textit{Same answer options as} 1
\end{enumerate}

\subsection{Post survey questionnaire}\label{A4}
\begin{itemize}
    \item P1: Which symptoms of discomfort you felt when using immersive devices in your configuration?
    \subitem $\square$~General discomfort \xspace$\square$~Fatigue \xspace$\square$~Headache \xspace$\square$~Eye strain \xspace$\square$~Difficulty focusing 
    
    \subitem $\square$~Salivation increasing \xspace$\square$~Sweating \xspace$\square$~Nausea \xspace$\square$~Difficulty concentrating \xspace$\square$~Fullness of the Head \subitem$\square$~Blurred vision \xspace$\square$~Dizziness \xspace$\square$~Vertigo \xspace$\square$~Stomach awareness \xspace$\square$~Burping \xspace$\square$~Other
    
    \item P2: For which category of users do you think the proposed system can be useful?
    \subitem \radiobutton All people \radiobutton People expert of the field \radiobutton People with disabilities \radiobutton I don't know \radiobutton Other
    
    \item P3: Do you think the proposed system was useful to use?
    \subitem Strongly disagree \radiobutton\radiobutton\radiobutton\radiobutton\radiobutton Strongly agree
    
    \item P4: Do you think the proposed system was interesting?
    \subitem \textit{Same answer options as} P3
    
    \item P5: Do you think the proposed system was easy to use?
    \subitem \textit{Same answer options as} P3

\end{itemize}

\end{document}